\documentclass[a4paper,fleqn]{cas-dc}
\usepackage[authoryear]{natbib}
\usepackage{amsmath}
\usepackage{graphicx}
\usepackage{booktabs}
\usepackage{array}
\usepackage{xcolor}
\newcolumntype{L}[1]{>{\raggedright\arraybackslash}p{#1}}

\newcommand{\bx}{\mathbf{x}}
\newcommand{\bX}{\mathbf{X}}
\newcommand{\bu}{\mathbf{u}}
\newcommand{\br}{\mathbf{r}}
\newcommand{\bF}{\mathbf{F}}
\newcommand{\bC}{\mathbf{C}}
\newcommand{\bS}{\mathbf{S}}
\newcommand{\bP}{\mathbf{P}}
\newcommand{\bK}{\mathbf{K}}
\newcommand{\bD}{\mathbf{D}}
\newcommand{\bz}{\mathbf{z}}
\newcommand{\bet}{\boldsymbol{\eta}}
\newcommand{\cW}{\mathcal{W}}
\newcommand{\dd}{\mathrm{d}}
\newcommand{\fx}[1]{\textcolor{black!45}{#1}}%

\begin{document}
\let\WriteBookmarks\relax
\def\floatpagepagefraction{1}
\def\textpagefraction{.001}

\shorttitle{Optimized size, shape and orientation of rhombic transposition flaps}
\shortauthors{N. Aage, B. M. Parks}

\title [mode = title]{Optimized size, shape and orientation of rhombic transposition flaps and excisions by
gradient-based methods}

\author[1]{Niels Aage}[orcid=0000-0002-3042-0036]
\cormark[1]
\ead{naage@dtu.dk}
\author[1]{Brandon M. Parks}
\affiliation[1]{organization={Technical University of Denmark, Department of Civil and Mechanical Engineering},
            addressline={Koppels All\'e},
            city={Kongens Lyngby},
            postcode={2800},
            country={Denmark}}
\cortext[1]{Corresponding author}

\begin{abstract}
This work presents a new method for gradient-based optimization of the rhombic transposition flap and the excision
it closes, which set the strain of the skin after closure. Previous studies swept at most three parameters. Here,
using structural optimization methods, up to eight variables, two flap angles, two cut lengths and the size, shape and
position of the rhombic excision, are optimized under up to 32 geometric and stress constraints. The skin is a
pre-stretched, incompressible hyperelastic membrane that wrinkles instead of carrying compression. Smoothed level-sets
on a fixed grid represent the cuts, and springs between interpolated points the sutures, so designs change without
remeshing and the adjoint method gives the sensitivities. The energy the surgery adds to the skin is minimized, and the
model reproduces the energies of a published 48-design sweep with a median deviation of 5\,\%. The optimal flap angles,
$(-23^\circ, 31^\circ)$, lie outside the swept range and add 63\,\% less energy than the Limberg flap and 15\,\% less
than the best swept design. Under unequal in-vivo tension, the optimal flap turns with the direction of the tension. Fitting the excision to the
lesion and its margin lowers the energy by a further 35\,\%, and with all eight variables it is 78\,\% below the
Limberg flap. This optimum is the same for finer grids, narrower cuts and wider undermining, although wider undermining
reduces its gain. Lowering its suture-line stress by 41\,\% costs only 2.3\,\% more energy.
\end{abstract}

\begin{graphicalabstract}
\includegraphics[width=\linewidth]{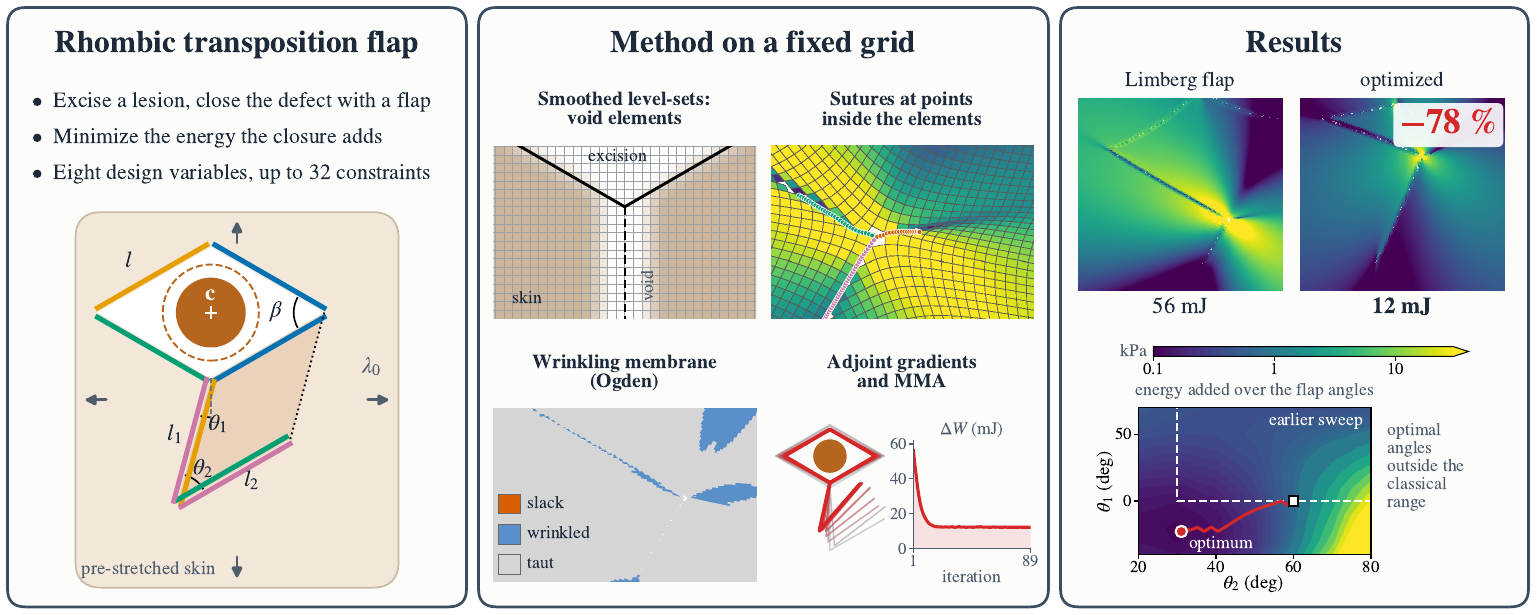}
\end{graphicalabstract}

\begin{highlights}
\item Flap and excision optimized with eight variables and exact gradients on a fixed grid
\item Optimal flap angles add 15\,\% less energy than the best of a 48-design sweep
\item Fitting the excision to the lesion margin is the largest single gain, 35\,\%
\item The optimal flap turns with the direction of the in-vivo skin tension
\item Lowering the suture-line stress by 41\,\% raises the energy by only 2.3\,\%
\end{highlights}


\maketitle

\section{Introduction}\label{sec:intro}

\begin{figure*}[t]
\centering
\includegraphics[width=\linewidth]{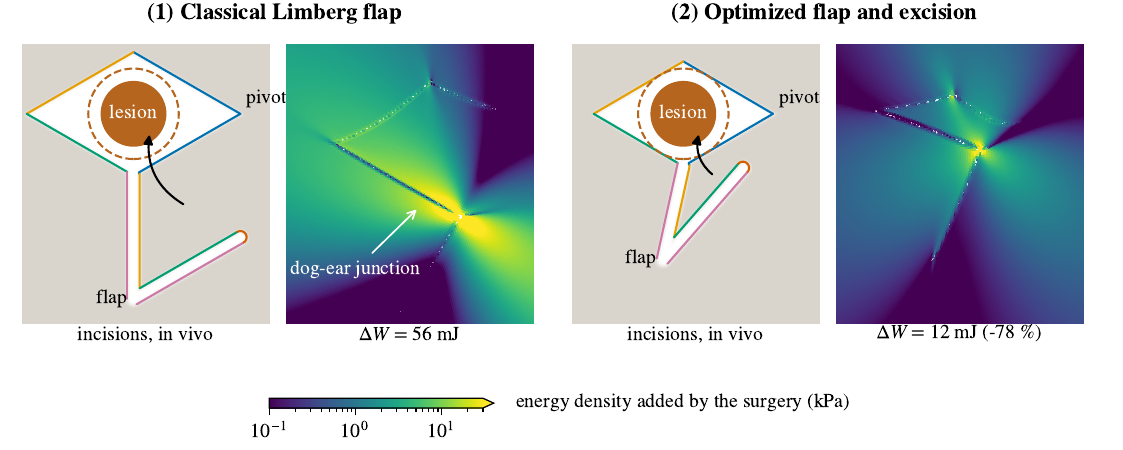}
\caption{The rhombic transposition flap. (1) The classical Limberg flap: a rhombic excision around the lesion and two
incisions that free a flap of the same shape, which is rotated about its pivot into the excision (arrow). Flap and excision follow a geometric rule, and the surgery
adds the most energy at the dog-ear junction of the closed wound. (2) The flap and the excision designed with all
eight geometric variables of the present work. The flap angles, the cut lengths and the excision are designed with
gradients, and the excision just covers the lesion margin. Left of each pair: the incisions in the pre-stretched
skin, with the lips that are sewn together colored by closure pair; right: the closed skin, colored by the energy
density added by the surgery (Section~\ref{sec:objective}) on a common logarithmic scale, with the energy $\Delta W$
the surgery adds to the skin.}
\label{fig:concept}
\end{figure*}

The excision of a skin lesion, most often a skin cancer, leaves a defect that is frequently too large to be closed
directly. Local flaps solve this by moving nearby skin into the defect, and the skin around the closed wound then
carries the tension of the closure. Excessive tension is regarded as a main cause of tissue necrosis, wound
dehiscence and hypertrophic scarring \citep{Paterno2011,Wong2012}. However, the geometry of a flap is still chosen
by geometric rules and the experience of the surgeon, and the tension that results is judged by touch
\citep{ListerGibson1972,Lee2019}. Thus, there is a strong potential for design methods that link the geometry of the
flap to the mechanical state of the closed skin. Fig.~\ref{fig:concept} illustrates this for the flap studied here:
designed with gradients, the flap and the excision add 78\,\% less strain energy to the skin than the classical
Limberg flap \citep{ListerGibson1972}.

The rhombic transposition flap (Fig.~\ref{fig:concept}) closes a rhombic defect with a flap of the same shape cut
next to it \citep{ListerGibson1972}. Two incisions free the flap, the flap is rotated about its base into the defect,
and its donor site is closed directly. In the classical design of Limberg, the rhombus has angles of $60^\circ$ and
$120^\circ$, the first incision extends the short diagonal of the rhombus, and the second incision is parallel to a
side of the rhombus. The variant of Dufourmentel changes these angles and fits rhombi with acute angles between
$60^\circ$ and $90^\circ$. Already \citet{ListerGibson1972}, who compared both designs in clinical practice,
concluded that predicting the behavior of these flaps would require a computer, as the designs were still partly
empirical. At the junction where the corner of the flap is sewn to the corner of the donor site, the closure gathers more skin
than can lie flat, and the excess rises out of the plane as a dog-ear. Both the tension of the closure and the dog-ear thus follow from the geometry of the flap,
which makes this geometry a natural subject for systematic design optimization.

Skin is a layered, nonlinear, anisotropic and viscoelastic tissue, and its properties vary between body sites and
individuals \citep{JoodakiPanzer2018}. For the simulation of wound closure it is commonly reduced to a single layer of
a hyperelastic material under the tension it carries in vivo. Skin cannot carry compression and wrinkles instead,
which tension-field membrane models capture by relaxing the strain energy \citep{MassaboGambarotta2007,Evans2009}. In
the present work the skin is an incompressible Ogden membrane with this relaxation, as in \citet{Spagnoli2022}. We
remark that the method proposed in this work is not tied to this constitutive model.

The mechanics of local flaps has been studied with the finite element (FE) method since the work of
\citet{DeHoff1981}. For the rhombic transposition flap, \citet{Topp2014} vary three geometric parameters one at a
time, the height of the defect, the angle of the first incision and the width of the flap, in a two-dimensional
hyperelastic model, and find the lowest closure stress and strain when the two incisions meet at $30^\circ$ instead of
the $60^\circ$ of the Limberg flap. \citet{Rajabi2015} move from the rhombic flap towards the Z-plasty with an
FE-based design. \citet{Lee2019} predict the stress profiles of advancement, rotation and transposition flaps for
different ages and defect sizes with Gaussian process surrogates of FE simulations, and \citet{Stowers2021} use such
surrogates to include the uncertainty of the material behavior in the design of reconstructive procedures. The work
closest to the present one is that of \citet{Spagnoli2022}, who simulate the Z-plasty, the multiple Z-plasty and the
rhombic transposition flap with an incompressible Ogden membrane that wrinkles instead of carrying compression, under
an in-vivo pre-stretch. For the rhombic flap they evaluate 48 combinations of the two flap angles, sampled every
$10^\circ$, by three measures: the strain energy of the skin, the stress along the suture line and the area of slack
skin. The best of their designs, $(\theta_1,\theta_2)=(0^\circ,30^\circ)$ in the notation of
Section~\ref{sec:geometry}, is the configuration that \citet{Topp2014} found best, and it lowers all three measures
against the Limberg flap.

However, the aforementioned studies only compare a predefined set of geometries, i.e.\ they rely on sweeps over a few
parameters rather than on systematic design optimization methodologies. The cost of such sweeps grows rapidly with
the number of parameters, so the size and shape of the excision, the cut lengths and the orientation of the flap
relative to the skin tension have not been designed together with the flap angles. Thus, to the best of the authors'
knowledge, the geometry of a local flap and of the excision it closes has not been optimized with gradients over a
continuous design space.

In this paper, the rhombic transposition flap and its excision are designed by gradient-based optimization, building on methods from structural optimization. One of the foundations of these methods is the
description of the design on a fixed structured mesh, in which the boundaries are immersed instead of meshed, so that no
remeshing is needed \citep{SigmundMaute2013,Andreasen2020}.
\citet{Holm2024} have shown that incisions and sutures can also be modeled on a fixed structured grid, which avoids
remeshing as the design changes, and recreated a Z-plasty study of \citet{alberini_computational_2021} with it. The
contributions of the present work are:
(i) a design parametrization in which the excision and the incisions are represented as smoothed level-sets \citep{Sethian2000} on the fixed grid and the sutures as springs between interpolated points, so that up to eight geometric design variables
change continuously and the sensitivities of the energy and of the stress constraint follow from the adjoint method;
(ii) a comparison with the 48-design study of \citet{Spagnoli2022} in their own measures;
(iii) optimal designs for four problems of increasing design freedom (flap angles, cut lengths, excision, all eight
variables), which quantify how much each design choice contributes;
(iv) the effect of the direction of an unequal in-vivo skin tension on the optimal flap; and
(v) a measure of the suture-line stress that is meaningful for stiff discrete sutures, together with a constraint on
it.

The rest of the paper is organized as follows. Section~\ref{sec:methods} presents the skin model, the design
variables and their representation on the fixed grid, the suture model, the objective, the optimization problems and
their sensitivities. Section~\ref{sec:results} compares the model with \citet{Spagnoli2022} and presents the optimized
designs. Section~\ref{sec:discussion} discusses the limitations and the clinical reading of the results, and
Section~\ref{sec:conclusions} concludes.

\begin{figure*}[t]
\centering
\includegraphics[width=\linewidth]{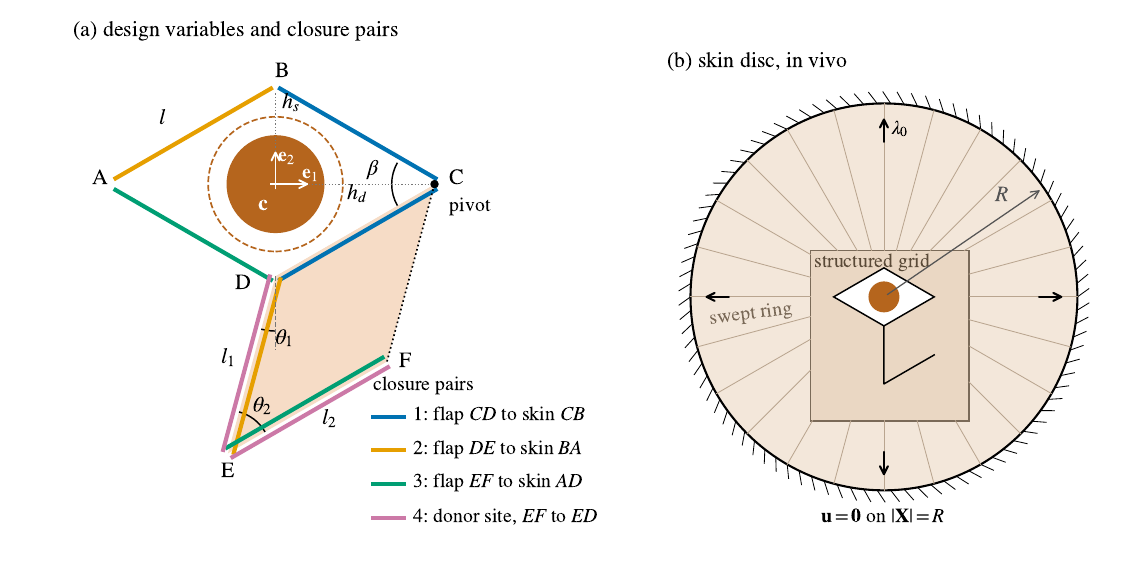}
\caption{Model problem. (a) Rhombic transposition flap with its design variables $l$, $\beta$, $c_x$, $c_y$,
$\theta_1$, $\theta_2$, $l_1$ and $l_2$, the unit vectors $\mathbf e_1$, $\mathbf e_2$ and the half diagonals $h_d$, $h_s$ of
Section~\ref{sec:geometry}, the lesion with its margin (dashed), the flap (shaded) with its uncut base
$FC$ (dotted), and the closure pairs 1 to 4 (pair 5 closes the round end of the band about $EF$,
Fig.~\ref{fig:embedding}(c)). Each lip is drawn on the side of the tissue it belongs to. (b) Skin disc of
radius $R$ with a fixed edge, pre-stretched in vivo by $\lambda_0$; a structured grid of square elements holds every
flap design, and a swept ring of elements connects it to the edge.}
\label{fig:problem}
\end{figure*}

\section{Materials and methods}\label{sec:methods}

\begin{figure*}[t]
\centering
\includegraphics[width=\linewidth]{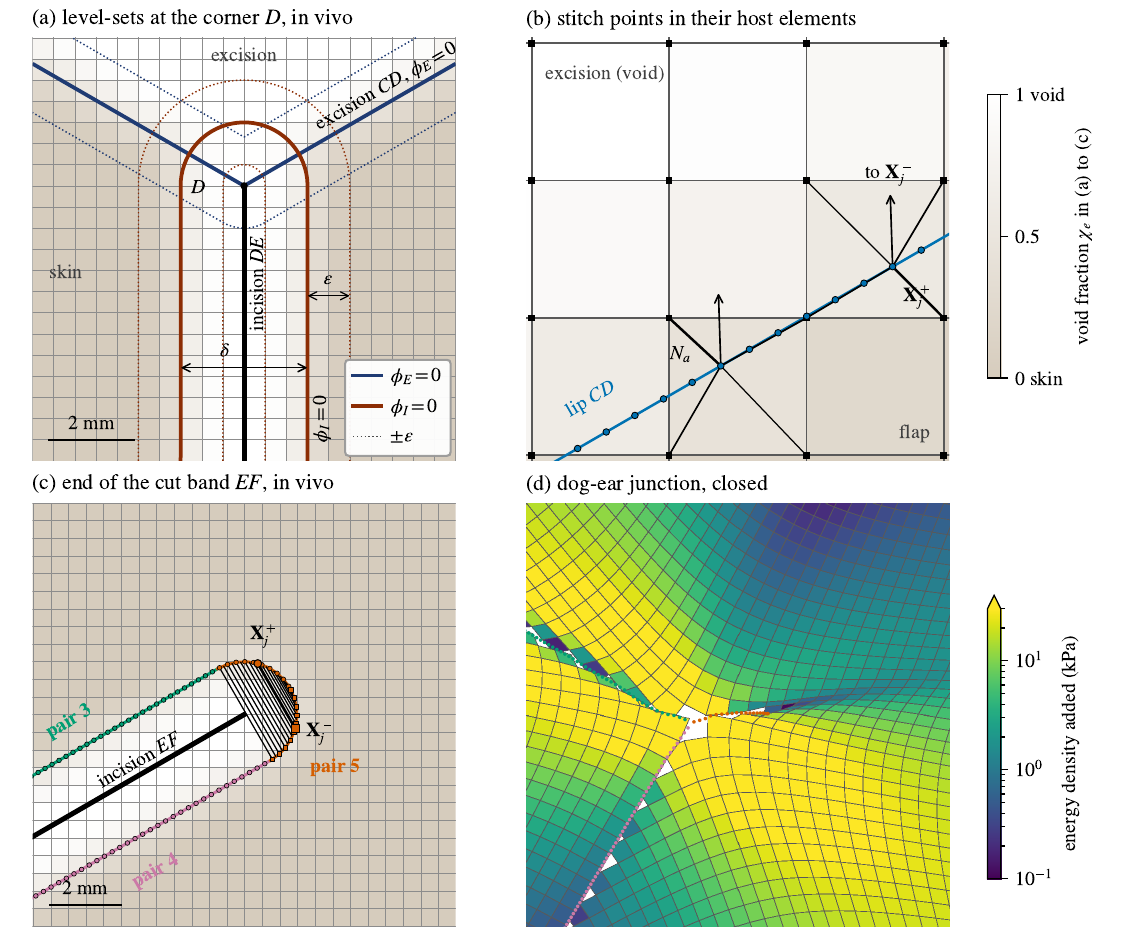}
\caption{Geometry and sutures on the fixed grid (Limberg flap, element size 0.5\,mm), in vivo in (a) to (c) and
closed in (d). (a) The first incision at the corner $D$ of the excision: void fraction $\chi_e$ of the elements (white: void), the excision line $CD$ and the incision line $DE$, the zero level-sets of $\phi_E$ and $\phi_I$ (solid) and the
edges of their smoothing zones at $\pm\varepsilon$ (dotted); $\phi_E=0$ lies on the excision line, and $\phi_I=0$ at
$\delta/2$ from the incision line.
(b) A few elements cut by a lip: the stitch points (circles), each joined to the four nodes of its host element, whose
shape functions interpolate its position and carry its spring force to these nodes, and the direction of its spring
(arrows), e.g.\ from $\bX_j^+$ to its partner $\bX_j^-$ on the other lip. (c) The end of the cut band about the incision line $EF$: the lips of
the closure pairs 3 (flap side), 4 (skin side) and 5 (round end), labeled in their colors, with every second stitch point, and the springs of pair 5 that
close the round end, one of them with its stitch points $\bX_j^+$ and $\bX_j^-$. (d) The dog-ear junction after closure, where the
flap corner $F$ has closed onto the donor-site corner $D$: the deformed elements, colored by the energy density added
by the surgery, and the closed seams.}
\label{fig:embedding}
\end{figure*}

Systematic gradient-based optimization needs a smooth and differentiable parametrization of the design variables.
Of equal importance, the mechanical equilibrium of the closed skin needs a robust solver, so that the optimizer does not
optimize model errors.
This section presents the model of the skin, the design variables and their representation on a fixed grid
(Figs.~\ref{fig:problem} and~\ref{fig:embedding}), the sutures that close the wound, the objective and the measures of the closed skin, as well as the optimization problem
and its sensitivities.

\subsection{Mechanical skin model}\label{sec:skin}
The mechanical problem is posed on a circular piece of skin, a disc of radius $R$ with a fixed edge, which is modeled as an
incompressible hyperelastic membrane of in-vivo thickness $t$. The disc is the undermined skin: the surgeon frees the skin
around the incisions from the subcutaneous tissue so that it can move in its plane, whereas the skin beyond stays
attached \citep{Spagnoli2022}. The edge of the disc is therefore fixed, and its radius is the extent of the
undermining. The computational grid is the in-vivo configuration. The tissue is stretched by $\bF_0$ before the
surgery, so the deformation gradient of the material is $\bF = \bF_n\bF_0$, where $\bF_n = \mathbf I + \nabla\bu$ is
the in-plane deformation caused by the surgery. The in-vivo stretch is either equibiaxial, $\bF_0 = \lambda_0\mathbf
I$, or unequal, with the stretch $\lambda_a$ along the direction $\mathbf m = (\cos a, \sin a)$ and $\lambda_b$ perpendicular to it, $\bF_0 = \lambda_a\,\mathbf m\otimes\mathbf m + \lambda_b\,\mathbf m_\perp\otimes\mathbf m_\perp$. In the model, the excised and cut tissue carries no in-vivo stress. On the fixed grid, it is represented by a weak
void material, as in immersed boundary and topology optimization \citep{SigmundMaute2013}
(Section~\ref{sec:geometry}).

With the principal stretches $\lambda_1\ge\lambda_2$ of $\bF$ and the thickness stretch $\lambda_3 =
1/(\lambda_1\lambda_2)$, the strain energy per unit volume used in this work is
\begin{equation}
  \begin{aligned}
  \cW &= \cW_{\mathrm O}(\lambda_1,\lambda_2) + \frac{\eta\,\mu_0}{2}\,(I_1-3),\\
  \cW_{\mathrm O}(\lambda_1,\lambda_2) &= \frac{\mu_1}{\alpha_1}\bigl(\lambda_1^{\alpha_1} + \lambda_2^{\alpha_1}
  + (\lambda_1\lambda_2)^{-\alpha_1} - 3\bigr),
  \end{aligned}
  \label{eq:ogden}
\end{equation}
where $\cW_{\mathrm O}$ is the one-term Ogden model \citep{Ogden1972} with the initial shear modulus $\mu_0 =
\mu_1\alpha_1/2$, $I_1 = \lambda_1^2 + \lambda_2^2 + (\lambda_1\lambda_2)^{-2}$, and $\eta = 0.01$. The second term is
added because of wrinkling. That is, a membrane cannot carry compression and wrinkles instead. Following the relaxed-energy
form of tension-field theory for biological membranes \citep{MassaboGambarotta2007}, the Ogden energy is relaxed in
three states: taut, where both principal stresses are tensile; wrinkled, where the membrane is in uniaxial tension
along the first principal direction and the energy is that of uniaxial tension, $\cW_{\mathrm
O}(\lambda_1,\lambda_1^{-1/2})$; and slack, where $\lambda_1\le1$ and the energy is zero. The relaxed energy is
continuously differentiable, which the Newton solver and the adjoint sensitivities rely on, but its tangent is singular
in two of the states. In wrinkled skin the energy depends on $\lambda_1$ alone, so there is no stiffness across the
wrinkles, and in slack skin it vanishes, so there is no stiffness at all. The small neo-Hookean term restores a small
stiffness in these regions for Newton's method. It does not prevent wrinkling, which the relaxation alone describes.

The membrane is discretized by finite elements in a total Lagrangian formulation on the in-vivo configuration, with
four-node bilinear elements and $2\times2$ Gauss points. The incompressibility only sets the thickness, $\lambda_3 = 1/(\lambda_1\lambda_2)$, at each point, and the in-plane
motion stays free. Hence, the displacement-based elements do not lock and need no mixed formulation. The relaxed energy, the stresses and the tangent are given in
Appendix~\ref{app:membrane}.

\subsection{Design variables of the rhombic transposition flap on a structured grid}\label{sec:geometry}
This section introduces the design variables of the rhombic transposition flap and how they are represented on a
fixed structured grid. First, the rhombus $ABCD$ is the excision, with side $l$, minor angle $\beta$ at $A$ and $C$,
and center $\mathbf c = (c_x, c_y)$, Fig.~\ref{fig:problem}(a). With $h_d = l\cos(\beta/2)$, $h_s = l\sin(\beta/2)$ and
the unit vectors $\mathbf e_1$ along the long diagonal and $\mathbf e_2$ normal to it,
\begin{equation}
  \begin{aligned}
  A &= \mathbf c - h_d\mathbf e_1,\qquad & C &= \mathbf c + h_d\mathbf e_1,\\
  B &= \mathbf c + h_s\mathbf e_2,\qquad & D &= \mathbf c - h_s\mathbf e_2 .
  \end{aligned}
\end{equation}
Next, the first incision starts at $D$, at the angle $\theta_1$ to the extension of the short diagonal $BD$ beyond $D$,
and the second starts at its end $E$, at the angle $\theta_2$ to $ED$:
\begin{equation}
  \begin{aligned}
  E &= D + l_1\,\mathbf R(-\theta_1)(-\mathbf e_2),\\
  F &= E + l_2\,\mathbf R(-\theta_2)\,\frac{D-E}{|D-E|} ,
  \end{aligned}
\end{equation}
with $\mathbf R$ the standard rotation matrix. The flap is $DEFC$, its base $FC$ is not cut, and it rotates about the pivot
$C$. A total of eight variables thus describe the excision and the flap, $l$, $\beta$, $c_x$, $c_y$, $\theta_1$,
$\theta_2$, $l_1$ and $l_2$. Two reference designs are used throughout: the Limberg flap, with $l=30$\,mm,
$\beta=60^\circ$, $\mathbf c=\mathbf 0$, $(\theta_1,\theta_2)=(0^\circ,60^\circ)$ and $l_1=l_2=l$, and the best design
of \citet{Topp2014} and \citet{Spagnoli2022}, which differs from it by $\theta_2=30^\circ$. Lastly, we require that the excision covers a circular lesion, which has a radius of 8\,mm at
the origin and a margin of 3\,mm (Fig.~\ref{fig:problem}(a)).

The design is represented on a fixed structured grid, over which the flap moves as the design changes. To this end, we extend the method of \citet{Holm2024}, which removes the elements of incisions and wounds from the grid
and joins groups of nodes by springs. Here, no element is removed, and the sutures attach at points inside the
elements. How this is done is detailed next. The basis of the method is a structured mesh in a box of square elements of
constant size $h$, which holds every design considered (Fig.~\ref{fig:problem}(b)).

We propose to represent the excision and the incisions by two level-sets \citep{Sethian2000}, evaluated at the element
centroids, one for each kind of cut. The excision covers an area and is described by the signed distance to the
rhombus $\mathcal R = ABCD$,
\begin{equation}
  \phi_E = \begin{cases}
  -d(\bX, \partial\mathcal R), & \bX\in\mathcal R,\\
  \hphantom{-}d(\bX, \partial\mathcal R), & \text{otherwise},
  \end{cases}
  \label{eq:phiE}
\end{equation}
with $d$ the distance to a set, here the boundary $\partial\mathcal R$ of the rhombus. The incisions are lines, but on a
fixed grid a cut of zero width cannot separate the elements. An incision therefore must cover a band of width $\delta$
about the polyline $DEF$, whose two segments share the vertex $E$, so that one level-set serves both incisions,
\begin{equation}
  \phi_I = \min\bigl(d(\bX, DE),\,d(\bX, EF)\bigr) - \delta/2 .
  \label{eq:phiI}
\end{equation}
Both level-sets are negative inside their cut, $\phi_E$ in the excision and $\phi_I$ in the band. The zero level-set
of $\phi_E$ therefore coincides with the excision line, whereas that of $\phi_I$ does not coincide with the incision
line: it lies at $\delta/2$ on either side of it, at the edges of the band (Fig.~\ref{fig:embedding}(a)). A sharp level-set would switch the elements it
crosses between skin and void in jumps as a cut moves over the grid. Smoothing it over the width $2\varepsilon$ instead
lets a cut move continuously over the grid, similar to level-set based structural optimization \citep{Andreasen2020}.
With the cubic smoothed Heaviside function
\begin{equation}
  H(x;\varepsilon) = \begin{cases}
  0, & x < -\varepsilon,\\
  \dfrac12 + \dfrac{3x}{4\varepsilon} - \dfrac{x^3}{4\varepsilon^3}, & |x| \le \varepsilon,\\
  1, & x > \varepsilon,
  \end{cases}
\end{equation}
the void fraction of element $e$ is
\begin{equation}
  \chi_e = 1 - \bigl(1-H(-\phi_E;\varepsilon)\bigr)\bigl(1-H(-\phi_I;\varepsilon)\bigr) .
  \label{eq:chi}
\end{equation}
Fig.~\ref{fig:embedding}(a) shows how the two level-sets map onto the grid at the corner $D$. The band must be at least $h+2\varepsilon$ wide, so that the element nearest to a cut is fully void and no element bridges the cut. The closure has to make up this band.

The smooth representation of the excision and the incisions requires special treatment of the weak material to avoid
a breakdown of the solver \citep{Wang2014}. To this end, intact elements, $\chi_e=0$, use the skin model of
Eq.~(\ref{eq:ogden}), with the element energy $\Psi_{T,e}(\bu_e) = t\int_{\Omega_e}\cW\,\dd A$ for the nodal
displacements $\bu_e$ of element $e$. Between the closing lips, the weak material is crushed, and with the nonlinear
law alone its elements would invert, with a negative Jacobian determinant. Fictitious domain and topology optimization
methods for finite strains meet the same problem in their void regions. \citet{Wang2014} therefore interpolate the
energy of such elements between the nonlinear model, evaluated at a scaled displacement, and a linear model, which
takes the rest of the deformation without such a limit. Here, the linear part is kept at the level of the void,
because a linear model is not rotation invariant: in the partly void elements along the lips of a flap turned by $60^\circ$, it would read a strain of 50\,\% and resist
the transposition. Void elements, $\chi_e=1$, thus use a weak linear material with a modulus $E_{\mathrm{void}}$ of
$10^{-6}$ times that of the skin, and partly void elements use the interpolation
\begin{equation}
  \Psi_e = (1-\chi_e)\,\Psi_{T,e}\bigl((1-\chi_e)\bu_e\bigr) + \chi_e\,\frac{E_{\mathrm{void}}}{E}\,\Psi_{L,e}(\bu_e),
  \label{eq:interp}
\end{equation}
where $\Psi_{L,e}$ is the element energy of the linear material with the small-strain response of the
skin.\footnote{Isotropic linear elasticity with the in-plane small-strain response of the incompressible membrane,
$\boldsymbol\sigma = 2\mu_0\bigl(\boldsymbol\varepsilon + (\mathrm{tr}\,\boldsymbol\varepsilon)\,\mathbf I\bigr)$,
whose uniaxial modulus is $E = 3\mu_0$. A change of area meets the finite stiffness $\mathrm{tr}\,\boldsymbol\sigma =
6\mu_0\,\mathrm{tr}\,\boldsymbol\varepsilon$, so this material does not lock either.} The skin part sees only $(1-\chi_e)\bu_e$, so its Jacobian determinant stays positive
while the lips crush the element. The linear part sees the full displacement, under which the element may turn inside
out, which a linear material tolerates, as it is evaluated in the undeformed geometry \citep{Wang2014}, and its energy is negligible.

\subsection{Suture model}\label{sec:sutures}
The purpose of the sutures is clearly to close the wound; on the fixed grid, however, they are the most delicate
part of the model. The lips, i.e.\ the
curves to which the sutures attach, lie on the zero level-sets of the excision and of the cut bands. This means that they will cross the elements and move through them as the
design changes. The sutures must therefore attach at points inside the elements, join two lips of different lengths,
and move continuously with the design, so that the energy of the closed skin stays differentiable
(Fig.~\ref{fig:embedding}(b)).

The closing mechanism of the problem at hand is predetermined. That is, the lips form five closure pairs
(Fig.~\ref{fig:problem}(a)): pairs 1 to 3 sew the flap edges $CD$, $DE$ and $EF$
to the skin edges $CB$, $BA$ and $AD$ of the excision, pair 4 closes the donor site by sewing the two sides of the band
$EF$ and $ED$ together, and pair 5 closes the round end of the band $EF$. This closure map is the same for every design; designing the map itself, e.g.\ by sewing other pairs of lips
together, is left for future work. At the corners, the lips are shaped so that every stitch holds tissue. The cut band
ends in half discs of radius $\delta/2$ at $D$ and $F$. At the flap tip $E$, the lips of the flap edges $DE$ and $EF$
are cut back to where they cross. At $D$, the skin lips along $AD$ and $ED$ end where their lines meet, outside the
half disc, and the flap lips along $CD$ and $DE$ follow the edge of the half disc to its midpoint if they meet at more
than $90^\circ$, and otherwise end where their lines meet. Pair 5 joins the two quarter circles of radius $\delta/2$
from the ends of the lips at $F$ to the tip of the band, with $\mathrm{round}(\pi\delta n_s/(4l_0))$ stitches on
each, 19 on the 0.5\,mm grid. Each lip of pairs 1 to 4 carries $n_s =
4l_0/h$ stitches at the fractions $s_j = (j-\tfrac12)/n_s$ of its arc length, with $l_0 = 30$\,mm the side of the
Limberg rhombus, so that every design has the same stitches. The two lips of a pair are joined in proportional arc length, inspired by the node-to-node method of
\citet{Spagnoli2022}, who follow the surgical practice of placing the stitches at proportional
distances from one end of the wound to the other. Numerically, a suture is a zero-length spring between the stitch points
$\bX_j^+$ and $\bX_j^-$, whose gap is interpolated with the shape functions $N_a$ of the elements that host the points (Fig.~\ref{fig:embedding}(b, c)),
\begin{equation}
  \mathbf t_j = \bx(\bX_j^+) - \bx(\bX_j^-) = \bigl[\bD(\bX+\bu)\bigr]_j,\quad
  \bx(\bX) = \sum_a N_a(\bX)\,\bx_a .
\end{equation}
The stitch points therefore move continuously with the design. The springs store
$\Pi_S = \tfrac12\sum_j\alpha_j|\mathbf t_j|^2$ with $\alpha_j = \kappa_s l_0/n_s$, the line stiffness $\kappa_s$
times the stitch spacing of the Limberg flap, and the shape functions that interpolate the position of a stitch point
also carry its spring force to the four nodes of its host element (Appendix~\ref{app:sutures}). The springs of a pair are thus
a quadrature, with one point per stitch, of a penalty on the gap along the lips. This is similar to yet another immersed boundary method from structural
optimization, namely to how interface integrals on cut elements are assembled. Here, Gauss points on the cut interface
are mapped into the parent element and weighted with its shape functions \citep{Dilgen2021}. However, two things differ. First, the two ends of a spring lie on
different lips, in different elements, so a spring ties two elements together. Second, the stitch points lie at fixed
fractions of the whole lip instead of at Gauss points of the pieces of the lip inside each element, which keeps the
two lips matched in proportional arc length and makes every stitch the same suture at every design. The quadrature is
then approximate, and its error changes as the design moves the lips over the grid. Our numerical studies show that four stitches per element length keep the energy smooth in the design. With one, stitches crossing element edges made the energy ripple by 0.6\,\%
(root mean square), with local minima every $3.4^\circ$ of $\theta_1$, and with four the ripple is 0.12\,\% on a 0.5\,mm grid.

The closure is thus a penalty, and $\kappa_s$ is a numerical parameter, which is calibrated once on the Limberg flap,
the design with the largest closing forces, as described next. A soft
$\kappa_s$ leaves the wound open, as a spring leaves a gap of its force over its stiffness. A stiff $\kappa_s$, however,
over-constrains the skin. Four stitches per element length tie two lips that cross the grid independently at more
points than their bilinear elements can follow, so the energy keeps rising with $\kappa_s$ instead of converging, an effect known as locking. This differs from \citet{Spagnoli2022}, who avoid both by meshing the two lips with matching nodes and tying these
node to node, which a fixed grid does not allow. With one constraint per element length of lip, i.e.\ one stitch per
element length or one spring on the mean gap of every four stitches, the energy converges as $1/\kappa_s$, and the two
closures reach nearly the same limit for rigid sutures, 0.2\,\% apart on the 0.5\,mm grid (Appendix~\ref{app:sutures}). They ripple
more with the design, however, so the optimization keeps a spring on every stitch and calibrates $\kappa_s$ against
this limit on the Limberg flap. With $\kappa_s = 10^4\,Et$ per millimeter
of seam, its energy is within 1\,\% of the limit, whereas $10^3\,Et$ leaves gaps of 0.9\,mm and an energy 7\,\% below this
limit, and $10^5\,Et$ gives an energy 2.4\,\% above it. Up to $10^6\,Et$, the energies of different designs change in
proportion within 5\,\%, and their ranking does not change. The large value of $\kappa_s$ reflects the steep
stiffening of the Ogden skin, which is stretched far into its stiffening range at the seams.

This leads to the final closed state $\bu$, which minimizes the total potential,
\begin{equation}
  \Pi(\bu) = \sum_e\Psi_e(\bu;\chi_e) + \frac{\Lambda}{2}\sum_j\alpha_j|\mathbf t_j|^2,\quad
  \br(\bu,\bz) = \frac{\partial\Pi}{\partial\bu} = \mathbf 0 ,
  \label{eq:state}
\end{equation}
and is reached by tightening the sutures, $\Lambda$ from $10^{-3}$ to 1 in growing steps (Appendix~\ref{app:sutures}). Each
step is solved by Newton's method with a line search on $\Pi$ and a sparse Cholesky factorization, shifted where the
tangent is not positive definite, to a relative residual of $10^{-9}$. Every closure ends at $\Lambda=1$, and the
stiff springs, not a constraint, close the wound. Unlike in \citet{Holm2024}, the tightening is not stopped at a gap
tolerance. With the same final stiffness for every design, the closed state minimizes the same potential and
depends smoothly on the design, which the adjoint sensitivities rely on, whereas a stopping rule on the gap would
change the final stiffness with the design in jumps. In the closed designs of this work, the largest gap that remains between two
stitch points is 0.05 to 0.29\,mm, at the ends of the cut bands and at the dog-ear junction, and the mean gap is 1 to
2.4\,$\mu$m, against gaps of up to 52\,mm before the closure. Fig.~\ref{fig:embedding}(d) shows the deformed grid at the
dog-ear junction of the closed Limberg flap.

\begin{figure*}[t]
\centering
\includegraphics[width=\linewidth]{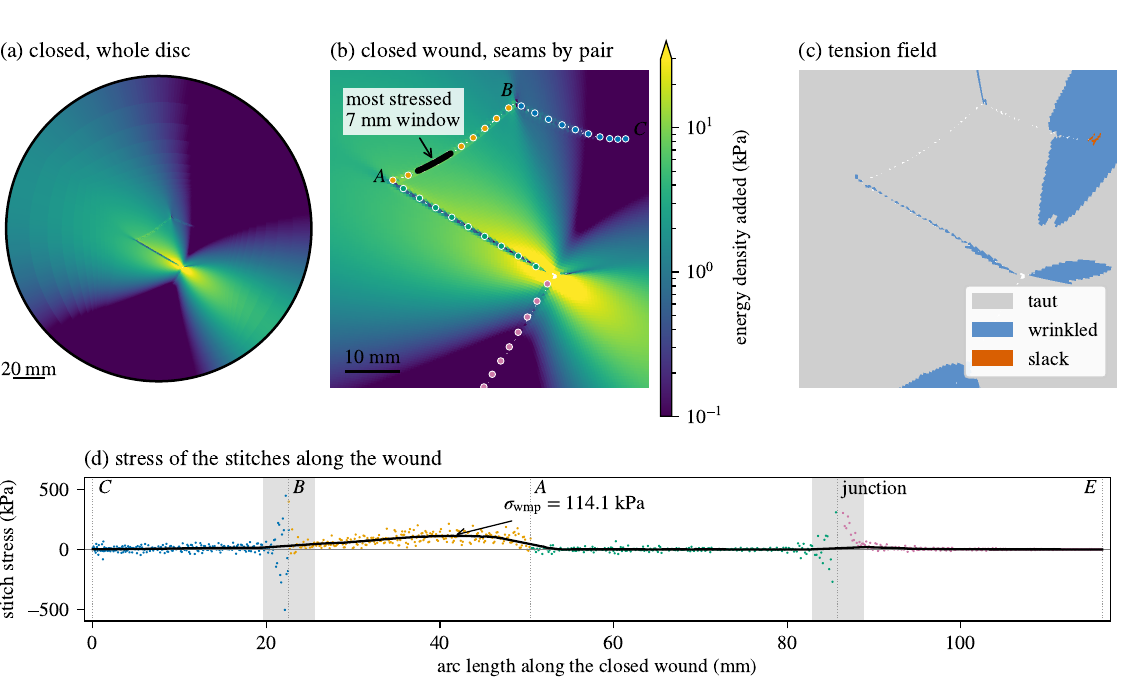}
\caption{The closed Limberg flap. (a) Energy density added by the surgery, $(\Psi_e(\bu)-\Psi_e(\mathbf 0))/(tA_e)$,
over the whole disc (logarithmic scale; values below 0.1\,kPa, including the slightly negative ones of relaxed skin,
in the lowest color); (b) the closed wound, with the seams marked every 3\,mm in the colors of their closure pairs (Fig.~\ref{fig:problem}(a)), the path of the suture-line stress from $C$ to $E$, and its most stressed 7\,mm window;
(c) taut, wrinkled and slack skin; (d) the stress $\sigma_j$ of every stitch along this path (dots) and its 7\,mm
mean (line), whose largest value is the suture-line stress $\sigma_{\mathrm{wmp}}$ (gray: the 3\,mm at the ends of
the cut bands that are left out). The energy concentrates at the dog-ear junction, where the flap corner $F$ closes
onto the donor-site corner $D$.}
\label{fig:model}
\end{figure*}

\subsection{Objective and measures of the closed skin}\label{sec:objective}
The objective is the energy that the surgery adds to the tissue that remains. It sums the strain that the closure
imposes, weighted by the stiffness of the skin, over all of the skin, and thus measures the closure tension that is
regarded as a main cause of complications (Section~\ref{sec:intro}),
\begin{equation}
  \begin{aligned}
  &\Delta W = W_{\mathrm{tot}} - W_{\mathrm{ref}},\\
  &W_{\mathrm{tot}} = \sum_{e:\,\chi_e<1}\Psi_e(\bu;\chi_e),\qquad
  W_{\mathrm{ref}} = \sum_{e:\,\chi_e<1}\Psi_e(\mathbf 0;\chi_e),
  \end{aligned}
  \label{eq:objective}
\end{equation}
where $W_{\mathrm{tot}}$ is the strain energy of the closed skin and $W_{\mathrm{ref}}$ that of the same tissue in the
in-vivo state; the void elements, whose weak material stores a negligible energy, are left out. As the skin is pre-stretched, $W_{\mathrm{tot}}$ also carries the in-vivo energy of the whole disc,
which grows with the area of the modeled skin and is 4.1\,mJ here. It stems from the in-vivo stretch alone: at
$\bu=\mathbf 0$, all tissue has the uniform energy density $\cW(\bF_0)=0.089$\,kPa, so $W_{\mathrm{ref}}$ vanishes for
$\bF_0=\mathbf I$. $\Delta W$ excludes this energy and thus measures
the closure itself. The excision changes $W_{\mathrm{ref}}$ by at most 0.03\,mJ between the designs of this work, so
minimizing $W_{\mathrm{tot}}$ would give nearly the same designs, but its constant part would dilute the relative
gains. Figs.~\ref{fig:concept}, \ref{fig:embedding}(d), \ref{fig:model}(a, b) and \ref{fig:designs} show the density
of $\Delta W$, the energy added per unit volume of in-vivo tissue, $(\Psi_e(\bu)-\Psi_e(\mathbf 0))/(tA_e)$ with the
in-vivo area $A_e$ of element $e$.

Two more measures describe the closed skin. The slack area is the area of skin in the slack state
(Fig.~\ref{fig:model}(c)), i.e.\ of skin that is not stretched in any direction relative to its natural state,
\begin{equation}
  A_s = \sum_{e\in\mathcal S} A_e ,\qquad
  \mathcal S = \bigl\{e : \chi_e<1/2,\ c_1(\bX_e)\le1\bigr\},
  \label{eq:slack}
\end{equation}
where $c_1(\bX_e)$ is the largest eigenvalue of $\bC$ at the centroid $\bX_e$ of element $e$
(Appendix~\ref{app:membrane}), so $\mathcal S$ holds the skin elements, $\chi_e<1/2$, that are slack at their
centroid. Such skin carries no stress and rises out of the plane, so its area measures the extent of dog-ears. The
suture-line stress
measures how much the sutures are loaded, as excessive tension of the closure is a main cause of wound dehiscence and
hypertrophic scarring (Section~\ref{sec:intro}). Here, the stress $\sigma_j$ that stitch $j$ transmits is the normal component
$f_{n,j}$ of its spring force over the current seam length $\Delta s_j$ it serves and the in-vivo thickness $t$,
positive in tension, and the suture-line stress is the largest mean of these stresses over 7\,mm of the closed wound,
\begin{equation}
  \sigma_j = \frac{f_{n,j}}{\Delta s_j\,t},\qquad
  \sigma_{\mathrm{wmp}} = \max_i\,\frac{1}{|\mathcal W_i|}\sum_{j\in\mathcal W_i}\sigma_j ,
  \label{eq:swmp}
\end{equation}
where the window $\mathcal W_i$ holds the stitches within 3.5\,mm of stitch $i$ along the closed wound, 7\,mm being the
length a single suture holds when sutures are placed 7\,mm apart. The stress $\sigma_j$ is thus a nominal normal
traction across the seam. The closed wound is taken as one path, from the pivot $C$ over $B$ and $A$ to the dog-ear
junction and along the donor-site seam to $E$
(Fig.~\ref{fig:model}(b, d)). The last 3\,mm of seam at the ends of the cut bands are left out, since real incisions
have no width and hence no such ends.

Two choices make this measure robust on the fixed grid. First, the stiff springs over-constrain the linear edges of
the elements, so the stitch forces alternate in sign from stitch to stitch about the net traction
(Fig.~\ref{fig:model}(d)). The mean of the signed $\sigma_j$ over 7\,mm is the net normal traction, whereas a mean of
$|\sigma_j|$ would add up the alternation. Second, the stress refers to the in-vivo thickness instead of the current
one. The stitch points lie in partly void elements that are crushed between the lips, and where a lip runs along
a grid line, a small change of the design moves them into the neighboring elements. At $\theta_1=0^\circ$, for
instance, the lips of the first incision run along grid lines, and turning it by $0.25^\circ$ changed a measure with the current thickness at the stitch points by 20\,\%.

The three measures $W_{\mathrm{tot}}$, $A_s$ and $\sigma_{\mathrm{wmp}}$ are those of the design study of
\citet{Spagnoli2022}, with which the model is compared in Section~\ref{sec:validation}. They use the slack area to
evaluate the extent of dog-ears, and $\sigma_{\mathrm{wmp}}$ is their wound mean peak stress.

\subsection{Optimization problems}\label{sec:problems}
With the closed state, the objective and the measures defined, the design problems can be stated. The state
equation~(\ref{eq:state}) is solved for every design, so the displacements depend on the design, $\bu=\bu(\bz)$, and the optimization problem is solved in a nested fashion. It reads
\begin{equation}
  \begin{aligned}
    \min_{\bz}\;:\;& \Delta W\bigl(\bu(\bz),\bz\bigr)/\Delta W_0 \\
    \text{s.t.}\;:\;& g_i\bigl(\bu(\bz),\bz\bigr) \le 0,\qquad i=1,\dots,m, \\
                    & \mathbf 0\le\bz\le\mathbf 1 ,
  \end{aligned}
  \label{eq:problem}
\end{equation}
where $\bz$ is the vector of the scaled design variables, $g_i$ are a group of geometric constraints, discussed next,
and the stress constraints, and $\Delta W_0$ is one hundredth of the energy added at the start design, so that the objective starts at 100. Each design variable
$v_k$ is scaled to $[0,1]$ by its bounds, which is necessary for a well-conditioned optimization problem:
$v_k = \underline v_k +
z_k(\overline v_k - \underline v_k)$, with $\theta_1\in[-40^\circ,70^\circ]$, $\theta_2\in[20^\circ,80^\circ]$,
$\beta\in[40^\circ,90^\circ]$, $l,l_1,l_2\in[20,40]$\,mm and $c_x,c_y\in[-10,10]$\,mm.

The geometric constraints are necessary in order to keep the design inside the grid and clinically sensible. They
consist of:
\begin{itemize}
\item Box: every vertex $A$ to $F$ lies at least 5\,mm inside the structured box, which gives four linear constraints
on the coordinates of each vertex, 24 in all (Appendix~\ref{app:parameters}).
\item Lesion: the lesion of radius $r_L$ and center $\mathbf c_L$ with its margin $m_L$ lies inside the rhombus, one
constraint per side, Eq.~(\ref{eq:lesion}).
\item Flap base: the uncut base of the flap is at least 15\,mm long, $g_5 = 1 - |F-C|/15\,\mathrm{mm}\le 0$.
\item Scar: the scar is not longer than that of the Limberg flap, $g_6 = (2l+l_1+l_2)/120\,\mathrm{mm} - 1\le0$.
\item Neck: the end $F$ of the second cut lies at least 4\,mm outside the side $CD$ of the excision,
$1-(F-C)\cdot\mathbf n_3/4\,\mathrm{mm}\le0$, with $\mathbf n_3$ the outward normal of $CD$. Within 2.5\,mm of $CD$, half the cut band plus the smoothing of the
excision, no tissue is left, so a closer $F$ would cut the flap off from its base.
\end{itemize}
With the outward normal $\mathbf n_k$ of side $k$ of the rhombus and its vertex $P_k$, the lesion constraints read
\begin{equation}
  g_k = \bigl((\mathbf c_L - P_k)\cdot\mathbf n_k + r_L + m_L\bigr)/l_0 \le 0,\qquad k=1,\dots,4,
  \label{eq:lesion}
\end{equation}
where the projection $(\mathbf c_L - P_k)\cdot\mathbf n_k$ is the signed distance of the lesion center from side $k$,
negative inside. The constraint is exact for the convex rhombus.

As the suture-line stress is of clinical interest (Section~\ref{sec:objective}), it can be constrained as well. The 7\,mm window means $s_i$ of the stitch stress $\sigma_j$, Eq.~(\ref{eq:swmp}), one centered at each stitch of pairs 1 to 4, are smoothed at the edges of the windows and of the left-out band ends over 0.5\,mm, so that each is
differentiable, and the largest of them lies 1 to 6\,\% below $\sigma_{\mathrm{wmp}}$. Their number, $4n_s$, grows as
the grid is refined. They are therefore aggregated into one constraint by the $p$-mean of their positive parts,
\begin{equation}
  \begin{aligned}
  g_7 &= \frac{\zeta_k\,\sigma_{\mathrm{PM}}}{f\bar\sigma} - 1 \le 0,\\
  \sigma_{\mathrm{PM}} &= \Bigl(\frac{1}{4n_s}\sum_{i=1}^{4n_s}\max(s_i,0)^p\Bigr)^{1/p},
  \end{aligned}
  \label{eq:pmean}
\end{equation}
which bounds the suture-line stress by a fraction $f$ of $\bar\sigma$, the largest window mean $s_i$ at
$(0^\circ,30^\circ)$. The positive part $\max(s_i,0)$ leaves out windows in net compression. It has a kink at
$s_i=0$, but its $p$-th power is $p-1$ times continuously differentiable, so $\sigma_{\mathrm{PM}}$ is differentiable as long
as one window is in tension, which holds at every design. With
$p=60$, the $p$-mean lies below the largest window mean, and the factor $\zeta_k$ corrects it towards that maximum by the
adaptive normalization of \citet{le_stress-based_2010}, as in \citet{daSilva2021}: $\zeta_k = q\,(\max_i
s_i/\sigma_{\mathrm{PM}})_{k-1} + (1-q)\,\zeta_{k-1}$ in iteration $k$, with $\zeta_1=1$ and $q=0.5$.

In all, the problems have up to eight design variables, up to 31 geometric constraints and one stress constraint.
Five problems of increasing complexity are solved, first to compare with the design study of \citet{Spagnoli2022} and
then to quantify the benefit of each added design freedom. In each, the variables that are not designed keep their
values of the Limberg flap:
\begin{itemize}
\item[P1] the flap angles $(\theta_1,\theta_2)$, started from the Limberg flap, with the box and neck constraints only;
\item[P2] the flap angles and the cut lengths $(\theta_1,\theta_2,l_1,l_2)$, started from $(0^\circ,30^\circ)$, with the
box and neck constraints only; without the scar bound, its optimum has a scar 2\,\% longer than that of the Limberg flap;
\item[P3] the excision and the flap angles $(l,\beta,c_x,c_y,\theta_1,\theta_2)$ with $l_1=l_2=l$, started from the
optimum of P1, with all 31 geometric constraints;
\item[P4] all eight variables, started from the Limberg flap and, as a second start, from $(0^\circ,30^\circ)$, with all
31 geometric constraints;
\item[P5] P4 with the stress constraint for $f=0.5$, 0.4 and 0.3, started from the Limberg flap.
\end{itemize}
The problems are solved by the Method of Moving Asymptotes (MMA) \citep{Svanberg1987} with an outer move limit of
0.03 on $\bz$. With this scale, the penalty $c$ of MMA on the constraint
violations is $10^5$, the usual value of $10^3$ multiplied by 100. A run stops when the largest change of a variable is below
$10^{-3}$ and the constraints hold to $10^{-3}$, or after 100 iterations (200 for P5).

\subsection{Sensitivity analysis}\label{sec:sensitivities}
Following standard adjoint sensitivity analysis \citep{Haftka1992}, based on the symmetric tangent stiffness $\bK$ of the closed state
and the adjoint $\bet$, the sensitivity of the objective can be computed as
\begin{equation}
  \frac{\dd\Delta W}{\dd z_k} = \frac{\partial\Delta W}{\partial z_k}
  + \bet^{\top}\frac{\partial\br}{\partial z_k},\qquad
  \bK\bet = -\frac{\partial\Delta W}{\partial\bu} .
\end{equation}
Only the end state enters the sensitivities. The closed state minimizes the potential~(\ref{eq:state}) at $\Lambda=1$
for every design, whatever the tightening path, so the path is not differentiated. The design enters
the residual through the void fractions $\chi_e$ and through the stitch points, which move with
the design at fixed nodal displacements. The explicit derivative of $\Delta W$ is that of the void fractions. The
derivatives of the vertices, the stitch points and the geometric constraints are computed by complex-step
differentiation \citep{SquireTrapp1998} and require no adjoints. The stress constraint of P5 takes one more adjoint solve with the same factorization of $\bK$, with
the right-hand side $-\partial g_7/\partial\bu$. Appendix~\ref{app:sensitivities} gives all terms and the
verification of the gradients against finite differences.

\section{Numerical examples}\label{sec:results}

To facilitate a direct comparison with results from the literature, the examples adopt the domain size, the lesion
position and the material parameters of \citet{Spagnoli2022}: a skin disc of radius $R=100$\,mm centered on the
lesion, with the Ogden parameters that they adopted from an in-vivo experiment on the forearm (Table~\ref{tab:params}
in Appendix~\ref{app:parameters}). The slope of the uniaxial stress-stretch curve of this skin grows by a factor of about 350
between its natural state and a stretch of 1.3. Unless otherwise stated, the in-vivo stretch is equibiaxial, $\lambda_0=1.1$.

The finite elements, the optimization and the sensitivities are implemented in MATLAB, and unless otherwise stated the
following mesh is used. The structured box, $[-38,44]\times[-64,24]$\,mm, holds every design considered with square
elements of size $h=0.5$\,mm (28\,864 elements), and 14 layers of elements, growing by a factor of 1.25, connect it to
the edge of the disc (9\,520 elements, Fig.~\ref{fig:problem}(b)). The grid has 76\,090 free degrees of freedom. The
cut bands are $\delta=3$\,mm wide and smoothed over $\varepsilon=1$\,mm, the narrowest band, $h+2\varepsilon$, on the
1\,mm grid of Fig.~\ref{fig:angles}(a), kept on the finer grids so that all grids model the same cuts, and every lip carries $n_s=240$ stitches. A
closure takes 35 to 55\,s with about 110 Newton iterations on two threads.

Section~\ref{sec:validation} first validates the model against the design study of \citet{Spagnoli2022}.
Sections~\ref{sec:angles} and~\ref{sec:freedom} then optimize the flap with increasing design freedom,
Section~\ref{sec:orientation} turns to the direction of the in-vivo tension and Section~\ref{sec:suture} to the stress
along the suture line, and Section~\ref{sec:mesh} checks the mesh, the size of the disc and the width of the cut
band.

\subsection{Comparison with the design study of Spagnoli et al.}\label{sec:validation}
\begin{figure*}[t]
\centering
\includegraphics[width=\linewidth]{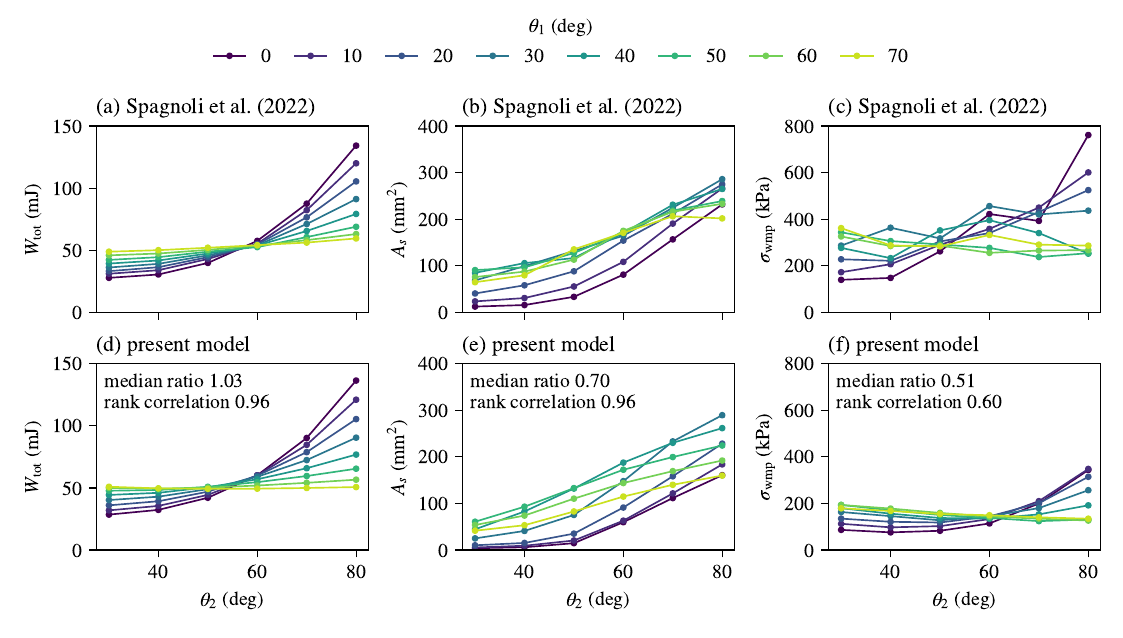}
\caption{The 48 designs of \citet{Spagnoli2022} in their three measures: strain energy $W_{\mathrm{tot}}$, slack area
$A_s$ and suture-line stress $\sigma_{\mathrm{wmp}}$, against $\theta_2$ for each $\theta_1$. Top: the values of
\citet{Spagnoli2022}, as plotted in their Fig.~15; bottom: the present model on the 0.5\,mm grid, with the median of the present values
over theirs and the rank correlation over the 48 designs.}
\label{fig:validation}
\end{figure*}

To validate the model and the optimization setup, the 48 designs of \citet{Spagnoli2022}, $\theta_1 =
0^\circ,\dots,70^\circ$ and $\theta_2 = 30^\circ,\dots,80^\circ$ in steps of $10^\circ$, were closed with the present
model and evaluated in their three measures, Fig.~\ref{fig:validation}. The strain energy is the measure that matters
most here: the objective, $\Delta W = W_{\mathrm{tot}} - W_{\mathrm{ref}}$, differs from $W_{\mathrm{tot}}$ only by
$W_{\mathrm{ref}}$, which changes by at most 0.03\,mJ between the designs, so an agreement in $W_{\mathrm{tot}}$
carries over to the objective. The energies agree well: $W_{\mathrm{tot}}$ deviates from their values by 5\,\% in the
median and by at most 15\,\% (median ratio 1.03), the designs are ranked alike (rank correlation 0.96), the energy curves cross near $\theta_2=60^\circ$ as
theirs, and within their range the best design is also $(0^\circ,30^\circ)$. Part of this agreement is due to the
3\,mm cut band, since with a band of 1.25\,mm the energies lie about a fifth below theirs (Section~\ref{sec:mesh}). The
other two measures are secondary checks. The slack area follows the same trends but is smaller. The suture-line
stress is about half of theirs and follows their trends less closely. Part of this difference is the thickness: with
the current instead of the in-vivo thickness at the stitch points, the median ratio to their values rises from 0.51
to 0.83 (Section~\ref{sec:objective}). The rest most likely comes from the stress peaks at the ends of the cuts near
the dog-ear junction, which the present measure leaves out, whereas the zero-width cuts of \citet{Spagnoli2022} have
no such ends. In summary, the model reproduces the strain energies of \citet{Spagnoli2022} with a median deviation of
5\,\% and ranks the 48 designs as they do, which validates it as the basis of the optimization.

\subsection{Flap angles}\label{sec:angles}
\begin{figure*}[t]
\centering
\includegraphics[width=\linewidth]{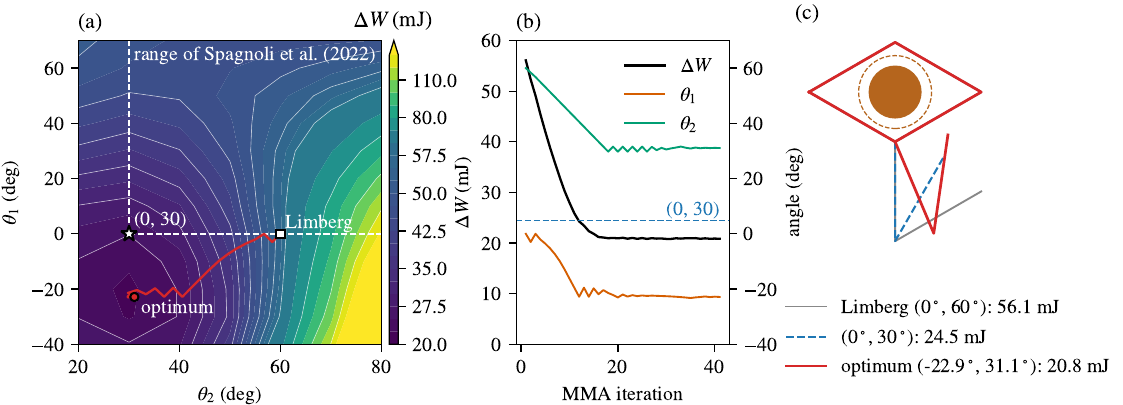}
\caption{Flap angles (problem P1). (a) Energy added by the surgery over the flap angles on the 1\,mm grid (84 designs,
contours), the range of the sweep of \citet{Spagnoli2022} (dashed), and the MMA path from the Limberg flap on the
0.5\,mm grid. (b) Energy added and the two angles against the MMA iteration. (c) Limberg flap, the best swept design
and the optimum, with the lesion and its margin.}
\label{fig:angles}
\end{figure*}

The sweep of \citet{Spagnoli2022} samples the flap angles in steps of $10^\circ$ and only for $\theta_1\ge0^\circ$.
Problem P1 frees both angles continuously and, started from the Limberg flap, reaches
$(\theta_1,\theta_2)=(-22.9^\circ,31.1^\circ)$ after 41 iterations, Fig.~\ref{fig:angles}. The main result is that the
optimal flap lies outside the range of the sweep: it is narrower than the classical designs, with a flap base of
16\,mm against 22\,mm for $(0^\circ,30^\circ)$ and 30\,mm for the Limberg flap, and it is turned towards the pivot $C$.
It adds 63\,\% less energy than the Limberg flap and 15\,\% less than the best swept design, $(0^\circ,30^\circ)$, and
it does so at the same suture-line stress and with less than half the slack area (Table~\ref{tab:optima}). With its
base about half as long as the flap, however, such a narrow flap may not supply sufficient blood flow
\citep{Spagnoli2022}. Its base of 16\,mm nevertheless satisfies the bound of 15\,mm imposed on the flap base in P3
to P5, which stands in for the blood supply.

The path of MMA follows the valley of the energy over the angles, Fig.~\ref{fig:angles}(a). The valley is narrow
along $\theta_2$, and the optima of P1 to P4 all have $\theta_2$ between $27^\circ$ and $32^\circ$
(Table~\ref{tab:optima}). Along $\theta_1$ it is flat: the energy varies by less than 1\,\% between
$\theta_1=-25^\circ$ and $-15^\circ$. Hence $\theta_2$ is the angle that matters, whereas $\theta_1$ can be chosen
within this range at almost no cost. Note that a run restarted at a converged design first steps along the ripple of
the fixed-grid objective, so all runs start from a classical design.

\begin{table*}[t]
\centering
\caption{Optimized designs (Ogden skin, 0.5\,mm grid). $\Delta W$: energy added by the surgery; $W_{\mathrm{tot}}$,
$A_s$ and $\sigma_{\mathrm{wmp}}$: the measures of \citet{Spagnoli2022}. Fitted: the reference design with its excision
fitted to the lesion margin, $l=25.4$\,mm. Values in gray are not design variables of the problem, and the center
of the excision stays at $\mathbf c=\mathbf 0$ in all optima.}
\label{tab:optima}
\small
\setlength{\tabcolsep}{4.5pt}
\begin{tabular*}{\textwidth}{@{\extracolsep{\fill}}lrrrrrrrrrrrr@{}}
\toprule
 & \multicolumn{6}{c}{Design variables} & \multicolumn{5}{c}{Closed state} & \\
\cmidrule(lr){2-7}\cmidrule(lr){8-12}
Design & $l$ & $\beta$ & $\theta_1$ & $\theta_2$ & $l_1$ & $l_2$ & $\Delta W$ & vs Limberg & $W_{\mathrm{tot}}$ & $A_s$
 & $\sigma_{\mathrm{wmp}}$ & Iterations \\
 & (mm) & ($^\circ$) & ($^\circ$) & ($^\circ$) & (mm) & (mm) & (mJ) & & (mJ) & (mm$^2$) & (kPa) & \\
\midrule
Limberg & 30 & 60 & 0 & 60 & 30 & 30 & 56.07 &  & 60.13 & 59.8 & 114.1  &  \\
\citet{Spagnoli2022} & 30 & 60 & 0 & 30 & 30 & 30 & 24.46 & $-56$\,\% & 28.52 & 4.0 & 86.6  &  \\
Limberg, fitted & 25.4 & 60 & 0 & 60 & 25.4 & 25.4 & 29.44 & $-48$\,\% & 33.53 & 31.0 & 72.8 &  \\
$(0, 30)$, fitted & 25.4 & 60 & 0 & 30 & 25.4 & 25.4 & 15.44 & $-72$\,\% & 19.54 & 3.8 & 64.1 &  \\
\midrule
P1 angles & \fx{30} & \fx{60} & $-22.9$ & 31.1 & \fx{30} & \fx{30} & 20.83 & $-63$\,\% & 24.90 & 1.8 & 86.4  & 41 \\
P2 cut lengths & \fx{30} & \fx{60} & $-2.7$ & 30.6 & 28.3 & 33.9 & 19.05 & $-66$\,\% & 23.12 & 1.5 & 77.3  & 38 \\
P3 excision & 25.6 & 59.2 & $-21.1$ & 27.7 & \fx{25.6} & \fx{25.6} & 13.51 & $-76$\,\% & 17.61 & 1.5 & 67.7  & 45 \\
P4 from Limberg & 25.2 & 60.6 & 12.4 & 28.7 & 23.3 & 29.6 & 12.13 & $-78$\,\% & 16.23 & 2.0 & 45.7  & 89 \\
P4 from $(0, 30)$ & 25.6 & 59.4 & 5.2 & 27.0 & 27.5 & 32.1 & 12.19 & $-78$\,\% & 16.28 & 2.8 & 46.8  & 49 \\
P5 stress bound, $f=0.3$ & 24.2 & 65.2 & 25.5 & 29.1 & 20.8 & 30.3 & 12.41 & $-78$\,\% & 16.51 & 3.0 & 27.0  & 127 \\
\bottomrule
\end{tabular*}
\end{table*}

\begin{figure*}[t]
\centering
\includegraphics[width=\linewidth]{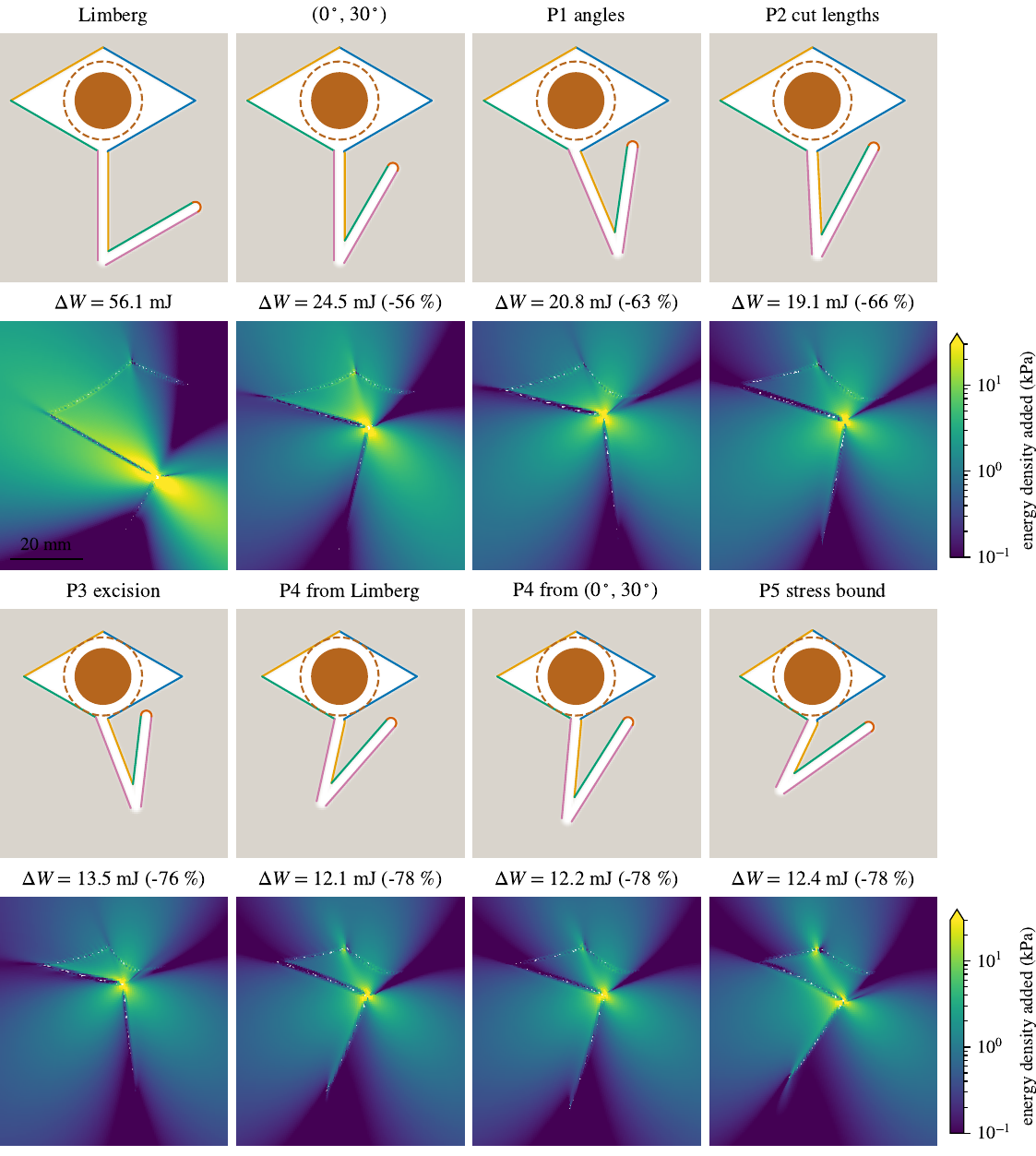}
\caption{Optimized designs of P1 to P5 (P4 from the Limberg flap and from $(0^\circ,30^\circ)$, P5 with the bound
$f=0.3$), with the Limberg flap and the best swept design for reference, in two blocks. In each block, top: the
excision, the lips of the closure pairs (colors as in Fig.~\ref{fig:problem}), the lesion and its margin, in the
in-vivo configuration; bottom: the closed states, colored by the energy density added by the surgery on a common
logarithmic scale, with the energy added and its change against the Limberg flap.}
\label{fig:designs}
\end{figure*}

\subsection{Maximum design freedom: cut lengths, excision and all eight variables}\label{sec:freedom}
The flap angles alone lower the energy added by the surgery by 63\,\%. Problems P2 to P4 free the remaining variables,
first the cut lengths, then the excision and finally all eight, and Table~\ref{tab:optima} and
Fig.~\ref{fig:designs} show what each added set of variables contributes. Free cut lengths (P2) add little: the energy
falls by 9\,\% against the optimum of P1, mainly through a longer second cut. P2 also has several local minima along
the flat valley in $\theta_1$. Started from the optimum of P1, MMA stays close to its start, and the flap of the
optimum of P4, placed on the Limberg rhombus, adds slightly less energy than the optimum of P2. In both, the second cut
is longer than the side of the rhombus.

The excision is the largest lever. With the size, shape and position of the rhombus free (P3), the rhombus shrinks
onto the lesion with its margin until all four lesion constraints are active, and the energy falls by a further
35\,\% against the optimum of P1. The rhombus stays centered at the lesion, and the flap base ends at its bound of
15\,mm. Note that the smaller excision removes less pre-stretched tissue. The objective $\Delta W$ does not reward
this, so the gain is the mechanical effect of the smaller defect.

With all eight variables free (P4), the energy falls to 78\,\% below the Limberg flap and 10\,\% below the optimum
of P3, which was started from the optimum of P1. MMA comes within 1\,\% of the optimum after 23 of its 89 iterations and then moves along the flat valley in
$\theta_1$. Again, the lesion constraints and the flap base are active, and the second cut is longer than the first.
In all optima of P3 and P4, the minor angle stays at $\beta=59^\circ$ to $62^\circ$ (Tables~\ref{tab:optima}
and~\ref{tab:fine}), near the $60^\circ$ of the Limberg flap, although a square would cover the lesion with its margin
with 13\,\% less excised area. The closure, not the excised area, thus sets the shape of the excision.
Started from $(0^\circ,30^\circ)$, P4 ends with 0.5\,\% more energy, nearly the same excision and the same suture-line
stress, but a different flap, with a smaller $\theta_1$ and longer cuts (Fig.~\ref{fig:designs}). Along the valley,
$\theta_1$ thus trades against the cut lengths at nearly equal energy: the optimization defines the excision well,
whereas the flap is not unique.

The gains against the Limberg flap include the size of its excision. With $l=30$\,mm, the rhombus of
\citet{Spagnoli2022} extends 2\,mm beyond the margin of the lesion. Fitted to the margin, $l=25.4$\,mm, the Limberg
flap adds 29.44\,mJ and $(0^\circ,30^\circ)$ 15.44\,mJ (Table~\ref{tab:optima}), and the optimum of P4 adds 59 and
21\,\% less energy than these.

\begin{figure*}[t]
\centering
\includegraphics[width=\linewidth]{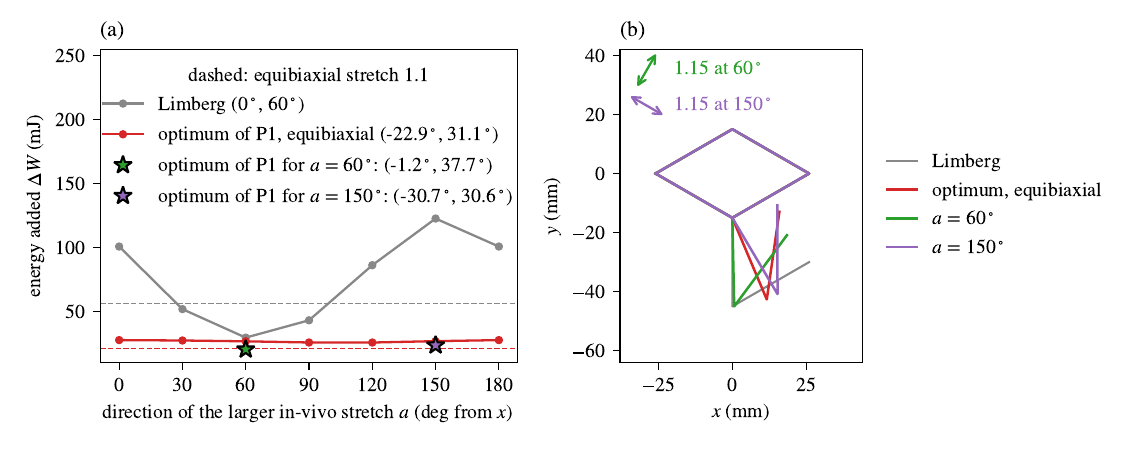}
\caption{Unequal in-vivo stretch, 1.15 along the angle $a$ from the $x$ axis and 1.05 across. (a) Energy added by the
Limberg flap and by the optimum for the equibiaxial stretch against $a$ (dashed: the equibiaxial values), and the
optima of P1 for $a=60^\circ$ and $150^\circ$ (stars; for $150^\circ$ the neck constraint is active). (b) These designs,
with the direction of the larger stretch.}
\label{fig:orient}
\end{figure*}

\subsection{Orientation to the in-vivo skin tension}\label{sec:orientation}
So far, the in-vivo stretch was equibiaxial. Skin, however, is pre-stretched more along some directions than
others. Scars that follow the lines of skin tension are the
least conspicuous \citep{Borges1960}, and the rhombic flap is planned so that the two corners that close its donor
site lie on the line of maximum extensibility of the skin, which usually runs at right angles to Langer's lines
\citep{ListerGibson1972}. With the unequal stretch $\lambda_a=1.15$ along the angle $a$ and $\lambda_b=1.05$ across it, whose area stretch
is within 0.2\,\% of the equibiaxial one,
the energy added by the Limberg flap varies fourfold with the direction $a$, Fig.~\ref{fig:orient}(a). It is least
where the closure of its donor site pulls across the larger stretch, i.e.\ along the more extensible direction, which
is the clinical rule. The optimum of P1 for the equibiaxial stretch, in contrast, adds nearly the same energy for every
direction, so it is far less sensitive to the orientation than the Limberg flap.

Solved for a given direction, P1 turns the flap with the tension, Fig.~\ref{fig:orient}(b). Adapted in this way, the
optimal flap adds 20.3 and 23.5\,mJ for the two directions computed, $a=60^\circ$ and $150^\circ$, where the Limberg
flap adds the least and the most energy, and 24 and 12\,\%
less than the equibiaxial optimum placed in the same tension. \citet{Topp2014} found the same tendency for the angle of
the first incision alone: the peak strain fell as the closure was aligned with the most extensible direction of the
skin. For $a=150^\circ$, the neck constraint is active, and the second cut ends at its bound, 4\,mm outside the side
$CD$ of the excision.

\begin{figure*}[t]
\centering
\includegraphics[width=\linewidth]{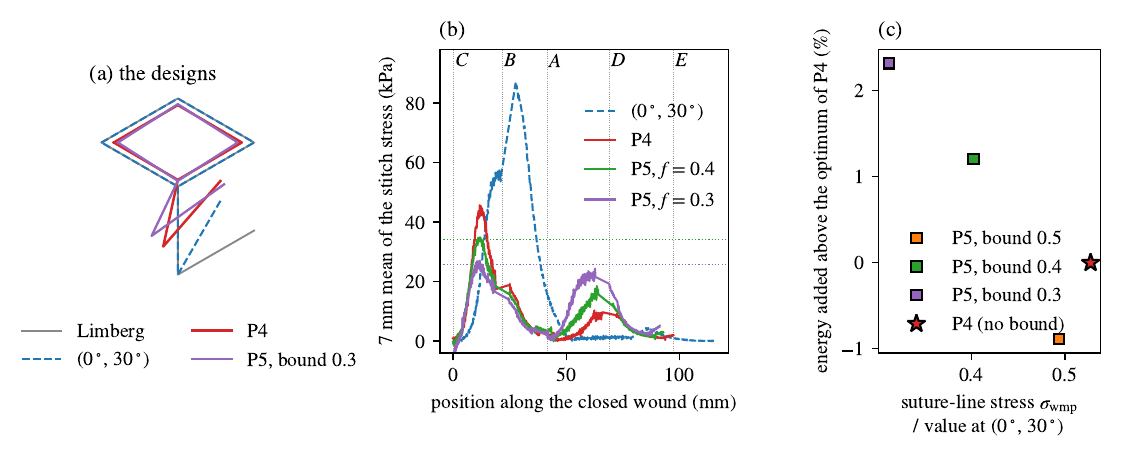}
\caption{Flap and excision under a bound on the suture-line stress (problem P5, all eight variables, $p$-mean bound).
(a) The Limberg flap, the best swept design, the optimum of P4 and the optimum of P5 with $f=0.3$. (b) The 7\,mm
mean of the stitch stress along the closed wound, from the pivot $C$ over $B$ and $A$ to the dog-ear junction $D$ and
along the donor-site seam to $E$ (seam ends of the optimum of P4 marked; 3\,mm at the ends of the cut bands left
out), for $(0^\circ,30^\circ)$, the optimum of P4 and the optima of P5 with $f=0.4$ and 0.3, with these bounds
$f\bar\sigma$ (dotted). The curves are not smoothed and peak up to 5\,\% above the smoothed bound. (c) Energy added
above the optimum of P4 against the suture-line stress reached, for $f=0.5$, 0.4 and 0.3.}
\label{fig:suture}
\end{figure*}

\subsection{Suture-line stress}\label{sec:suture}
The optima so far minimize the energy added by the surgery alone. A design that closes with less energy could still
load its sutures more, so this section first measures the suture-line stress of the optima and then bounds it. The
optimum of P1 has the same suture-line stress as $(0^\circ,30^\circ)$, and the optima of P2 to P4 lower it, by up to
47\,\% (Table~\ref{tab:optima}). Hence the energy optimum does not necessarily buy its lower energy with a higher
suture tension. The most stressed window, however, moves between the designs, Fig.~\ref{fig:suture}(b): it lies near
the corner $B$ at $(0^\circ,30^\circ)$, and in the middle of the seam $CB$ at the optima, where the flap edge $CD$ is
sewn to the excision. Wherever it lies, this peak loads the sutures, and a mean along the whole wound would hide it.
Although the analysis does not point to a higher stress for a lower energy, controlling the peak is still of
interest, as excessive tension of the closure is a main cause of dehiscence and hypertrophic scarring
(Section~\ref{sec:intro}).

Problem P5 bounds this peak by the $p$-mean of the window means, Eq.~(\ref{eq:pmean}), with the bounds
$f\bar\sigma=43$, 34 and 26\,kPa, corresponding to $f=0.5$, 0.4 and 0.3, below the 46\,kPa of the optimum of P4. To
lower the peak on $CB$, the flap turns further away from its pivot, to $\theta_1=18^\circ$ to $25^\circ$, and its first cut shortens,
while a second peak grows on the seam $AD$, Fig.~\ref{fig:suture}(a, b). The excision stays fitted to the lesion with
its margin, and the flap base at its bound. The price in energy is small: lowering the suture-line stress of the
energy optimum by a quarter costs about 1\,\% of energy, and by two fifths about 2\,\%, Fig.~\ref{fig:suture}(c) and
Table~\ref{tab:optima}.

With $f=0.5$, the bound is inactive at the end: the run moves along the flat valley in $\theta_1$ and ends with
0.9\,\% less energy and 6\,\% less stress than the optimum of P4. Measured from this better design, the bound
$f=0.3$ lowers the suture-line stress by 37\,\% for 3.2\,\% more energy. The valley thus holds designs of nearly equal energy
but different suture-line stress, and the bound selects among them, which is why the stress can be lowered so
cheaply. The run with $f=0.3$ is an outlier in its length: it needed 127 iterations to meet the bound, as the
correction $\zeta_k$ lags behind the peak that moves with the design. It and the run with $f=0.5$, with 113 iterations,
are the only runs of this work that took more than 100 iterations. Note that an iteration of P5 takes about as long
as one of P4, since the closure dominates, so the constraint adds no measurable cost.

\begin{table*}[t]
\centering
\caption{Optima of the flap angles (P1) and of all eight variables (P4) on the 0.25\,mm grid with cut bands of
$\delta=3$ and 1.25\,mm, and of P4 on discs of $R=200$ and 300\,mm, with the optima of the 0.5\,mm grid on the
100\,mm disc (Table~\ref{tab:optima}) for comparison. The rows without iterations are closed, not optimized, on the grid, band
and disc of their block: the Limberg flap, $(0^\circ,30^\circ)$ and the optima of another grid, band or disc. vs Limberg: change of $\Delta W$ against the Limberg flap of
the same block. Values in gray are not design variables of the problem, and the center of the excision stays at
$\mathbf c=\mathbf 0$ in all optima.}
\label{tab:fine}
\small
\begin{tabular*}{\textwidth}{@{\extracolsep{\fill}}lrrrrrrrrrr@{}}
\toprule
 & \multicolumn{6}{c}{Design variables} & \multicolumn{3}{c}{Closed state} & \\
\cmidrule(lr){2-7}\cmidrule(lr){8-10}
Design & $l$ & $\beta$ & $\theta_1$ & $\theta_2$ & $l_1$ & $l_2$ & $\Delta W$ & vs Limberg & $\sigma_{\mathrm{wmp}}$
 & Iterations \\
 & (mm) & ($^\circ$) & ($^\circ$) & ($^\circ$) & (mm) & (mm) & (mJ) & & (kPa) & \\
\midrule[\heavyrulewidth]
\multicolumn{11}{@{}l}{\textit{Reference}: grid $h=0.5$\,mm, band $\delta=3$\,mm, disc $R=100$\,mm (Table~\ref{tab:optima})} \\
\cmidrule{1-11}
Limberg & 30 & 60 & 0 & 60 & 30 & 30 & 56.07 &  & 114.1 &  \\
$(0, 30)$ & 30 & 60 & 0 & 30 & 30 & 30 & 24.46 & $-56$\,\% & 86.6 &  \\
P1 from Limberg & \fx{30} & \fx{60} & $-22.9$ & 31.1 & \fx{30} & \fx{30} & 20.83 & $-63$\,\% & 86.4 & 41 \\
P4 from Limberg & 25.2 & 60.6 & 12.4 & 28.7 & 23.3 & 29.6 & 12.13 & $-78$\,\% & 45.7 & 89 \\
\midrule[\heavyrulewidth]
\multicolumn{11}{@{}l}{\textit{Finer grid}: grid $h=0.25$\,mm, band $\delta=3$\,mm, disc $R=100$\,mm} \\
\cmidrule{1-11}
Limberg & 30 & 60 & 0 & 60 & 30 & 30 & 55.43 &  & 111.6 &  \\
$(0, 30)$ & 30 & 60 & 0 & 30 & 30 & 30 & 24.09 & $-57$\,\% & 86.1 &  \\
P1 from $(0, 30)$ & \fx{30} & \fx{60} & $-23.4$ & 30.8 & \fx{30} & \fx{30} & 20.49 & $-63$\,\% & 84.1 & 26 \\
P4 from Limberg & 25.2 & 60.9 & 18.3 & 27.5 & 22.8 & 30.4 & 11.92 & $-78$\,\% & 40.7 & 70 \\
P4 from $(0, 30)$ & 25.6 & 59.4 & 5.6 & 22.7 & 28.3 & 33.0 & 11.88 & $-79$\,\% & 62.5 & 48 \\
optimum of the 0.5\,mm grid & 25.2 & 60.6 & 12.4 & 28.7 & 23.3 & 29.6 & 11.99 & $-78$\,\% & 43.9 &  \\
\midrule[\heavyrulewidth]
\multicolumn{11}{@{}l}{\textit{Larger disc}: grid $h=0.5$\,mm, band $\delta=3$\,mm, disc $R=200$\,mm} \\
\cmidrule{1-11}
Limberg & 30 & 60 & 0 & 60 & 30 & 30 & 20.72 &  & 25.6 &  \\
$(0, 30)$ & 30 & 60 & 0 & 30 & 30 & 30 & 14.06 & $-32$\,\% & 31.2 &  \\
P4 from Limberg & 25.0 & 61.7 & 6.6 & 30.7 & 24.9 & 29.7 & 8.84 & $-57$\,\% & 24.1 & 81 \\
optimum of the 100\,mm disc & 25.2 & 60.6 & 12.4 & 28.7 & 23.3 & 29.6 & 8.71 & $-58$\,\% & 27.1 &  \\
\midrule[\heavyrulewidth]
\multicolumn{11}{@{}l}{\textit{Larger disc}: grid $h=0.5$\,mm, band $\delta=3$\,mm, disc $R=300$\,mm} \\
\cmidrule{1-11}
Limberg & 30 & 60 & 0 & 60 & 30 & 30 & 17.23 &  & 18.2 &  \\
$(0, 30)$ & 30 & 60 & 0 & 30 & 30 & 30 & 12.71 & $-26$\,\% & 25.0 &  \\
P4 from Limberg & 25.2 & 60.9 & 13.0 & 27.9 & 25.4 & 31.0 & 8.23 & $-52$\,\% & 21.6 & 84 \\
optimum of the 100\,mm disc & 25.2 & 60.6 & 12.4 & 28.7 & 23.3 & 29.6 & 8.21 & $-52$\,\% & 24.6 &  \\
\midrule[\heavyrulewidth]
\multicolumn{11}{@{}l}{\textit{Narrower band}: grid $h=0.25$\,mm, band $\delta=1.25$\,mm, disc $R=100$\,mm} \\
\cmidrule{1-11}
Limberg & 30 & 60 & 0 & 60 & 30 & 30 & 44.07 &  & 83.7 &  \\
$(0, 30)$ & 30 & 60 & 0 & 30 & 30 & 30 & 18.14 & $-59$\,\% & 63.0 &  \\
P1 from $(0, 30)$ & \fx{30} & \fx{60} & $-17.3$ & 31.0 & \fx{30} & \fx{30} & 16.39 & $-63$\,\% & 60.9 & 43 \\
P4 from Limberg & 25.0 & 61.5 & 3.7 & 30.8 & 23.9 & 28.4 & 9.62 & $-78$\,\% & 37.2 & 65 \\
P4 from $(0, 30)$ & 25.1 & 61.3 & 3.8 & 26.3 & 27.6 & 31.5 & 9.54 & $-78$\,\% & 39.8 & 39 \\
optimum of the 3\,mm band & 25.6 & 59.4 & 5.6 & 22.7 & 28.3 & 33.0 & 9.62 & $-78$\,\% & 53.9 &  \\
\bottomrule
\end{tabular*}
\end{table*}

\subsection{Mesh resolution, domain size and cut width}\label{sec:mesh}
The element size, the size of the disc and the width of the cut band are choices of the model, and this section
examines how they affect the results. Table~\ref{tab:fine} collects the optima of all three checks.

On a grid of half the element size, $h=0.25$\,mm (139\,936 elements, $n_s=480$, the same $\kappa_s$), the energies are 1 to 2\,\% lower, and the
ratios between designs agree within 0.3\,\%. Re-optimized on this grid from $(0^\circ,30^\circ)$, the optimum of P1 moves by about $0.5^\circ$,
and the two runs of P4, from the Limberg flap and from $(0^\circ,30^\circ)$, end at different points of the flat
valley, with nearly the same excision and the same active constraints as on the 0.5\,mm grid. The optimum of the
0.5\,mm grid, closed on the finer grid, adds less than 1\,\% more energy than the better of them. The 0.5\,mm grid
thus suffices for the optimization.

The disc is the undermined skin (Section~\ref{sec:skin}), so a larger disc means a wider undermining. On discs of
$R=200$ and 300\,mm, with $14(R-50)/50$, i.e.\ 42 and 70, ring layers, the energies of all designs fall, and most so for the Limberg flap, whose energy falls to about a
third. Re-optimized on these discs, P4 ends with nearly the same excision and a flap on the same flat valley, and the
optimum of the 100\,mm disc, closed on the larger discs, adds even slightly less energy than these runs reach. The
optimal design thus does not depend on $R$. Its gain against the Limberg flap, however, shrinks from 78\,\% to 52\,\% at $R=300$\,mm, since the Limberg flap
profits most from the wider undermining. The suture-line stresses fall as well, and their order changes: at
$R=300$\,mm, the Limberg flap has the lowest. The stress results of Section~\ref{sec:suture} therefore hold for the
100\,mm disc.

The cut band matters more than the grid. A band of $\delta=1.25$\,mm, the narrowest that keeps the element nearest to
a cut void on the finer grid ($\varepsilon=0.5$\,mm), lowers the energy of every design by about a fifth. The designs
keep their order, and the optima of P1 and P4 keep their gains against the Limberg flap, 63 and 78\,\%, and only move
along the flat valley. The gain of the optimum of P1 against $(0^\circ,30^\circ)$, however, shrinks from 15 to 10\,\%.

In summary, the grid, the disc and the cut band set the absolute energies, but none of them changes the optimized
excision, and the flap moves only along the flat valley of nearly equal energy. The gains against the Limberg flap do
not depend on the grid or the band, whereas they shrink with the extent of the undermining.

\section{Discussion}\label{sec:discussion}

This section assesses the findings as a whole, starting with the simplifications of the model that bound them. First, the
skin is a planar membrane, so the dog-ear cannot rise out of the plane. The excess tissue at the junction is instead
compressed in the plane and relaxed by wrinkling, and a shell model would be needed to resolve it. Second, the skin is
isotropic, with the parameters that \citet{Spagnoli2022} identified for the forearm, and anisotropy enters only
through the in-vivo stretch. Skin is anisotropic, and on the forearm of one subject, its stiffness along Langer's
line was about three times that across it \citep{Elouneg2023}. Its natural tension varies with the body site and age
\citep{Ayadh2023}, and the incision line with the lowest tension across the wound is not settled even for the forearm
\citep{Rosicka2021}. Third, only one lesion, of radius 8\,mm with a margin of 3\,mm, is considered.
Fourth, the closure map, i.e.\ which point of a lip is sewn to which, is fixed to proportional arc length, and the
sutures are a continuum of stiff springs. Hence the net traction over a few millimeters is meaningful, but the force of a single
stitch is not. The springs over-constrain the elements along the lips, so their stiffness is calibrated against a
closure without over-constraint instead of being raised until the energy converges (Appendix~\ref{app:sutures}). A tie of the
lips that neither over-constrains the grid nor ripples with the design would make this calibration unnecessary.
Fifth, the cuts are bands of 3\,mm, and the disc of undermined skin has a fixed size. A narrower band lowers the energy
of every design by about a fifth but keeps their order and the gain of the optimum against the Limberg flap, whereas a
wider undermining keeps the optimum but reduces its gain, from 78 to 52\,\% at $R=300$\,mm
(Section~\ref{sec:mesh}). The narrower band, however, shrinks the gain of the optimal angles over $(0^\circ,30^\circ)$
from 15 to 10\,\%, and on the 300\,mm disc the Limberg flap has the lowest suture-line stress. Sixth, the blood supply of the
flap is not modeled; the bound of 15\,mm on its base stands in for it, and the optima of P3 to P5 reach this bound.
The optimum of P1, which has no bound on its base, is nearly twice as long as its base of 16\,mm, and such narrow
flaps may not supply sufficient blood flow \citep{Spagnoli2022}.
The neck constraint keeps the flap joined to its base on the 0.5\,mm grid with the 3\,mm band. Its bound of 4\,mm
leaves at least 1.5\,mm, three elements, of intact skin between the cut band and the excision. It is a modeling
choice and not a clinical rule. At the optima, the end of the second cut lies 4.7 to 14\,mm outside the
excision, against 26\,mm for the Limberg flap, and the constraint is active only for $a=150^\circ$
(Section~\ref{sec:orientation}). Finally, healing and
growth are not modeled, and MMA, as any other gradient-based method, only finds local optima. Note, however, that a
sweep at the resolution of \citet{Spagnoli2022}, with six to eight values per variable, would need millions of
closures for eight variables and 32 constraints, whereas MMA solved every problem of this work within 100
iterations, except for two runs of P5, and often in far fewer.

Apart from these simplifications, the results suggest three rules for the rhombic transposition flap. First, the
excision should be no larger than the lesion with its margin requires. This is the largest gain, 35\,\% of the energy
added beyond the optimal angles. Second, the flap should be narrow, with $\theta_2$ of $23^\circ$ to $31^\circ$ and its base near the
lower bound of 15\,mm, whereas $\theta_1$ matters little, as the energy is nearly flat along it. With both cuts as long as the
side of the rhombus (P1, P3), the optimal flap is turned towards its pivot, $\theta_1=-23^\circ$ to $-17^\circ$; with
free cut lengths (P2, P4, P5), the optima lie at $\theta_1=-3^\circ$ to $25^\circ$, as $\theta_1$ trades against the
cut lengths at nearly equal energy. Third, under an unequal skin
tension the flap should be oriented so that the donor site closes along the more extensible direction, and its angles
should follow the tension. The first two rules extend the finding of \citet{Topp2014} and \citet{Spagnoli2022} that
$\theta_2=30^\circ$ is better than the classical $60^\circ$, and the third recovers the clinical rule of placing the
corners that close the donor site on the line of maximum extensibility of the skin \citep{ListerGibson1972}.

The energy added by the surgery was chosen as the objective for two reasons. First, it measures the total mechanical
cost of the closure for the tissue that remains. Second, it is continuously differentiable in the design, because the
relaxed membrane energy is. A strain measure of the closed skin was also tested, the $p$-mean of the relaxed von Mises strain, which
penalizes local strain peaks. With the Ogden skin, about 1\,000 to 1\,300\,mm$^2$ of skin wrinkle, the relaxed strain
is only continuous at the boundary between taut and wrinkled skin, and MMA did not converge on the strain measure
within 60 iterations, whereas the energy optimum also had the lower strain measure. Local limits are therefore better
imposed as constraints, as the suture-line stress is in P5.

The fixed grid is what makes the design space continuous, and hence, what enables the efficient optimization of the
plasty. Every design is closed on the same grid, the design variables move the level-sets and the stitch points
continuously, and the sensitivities follow from the adjoint method with the factorization of the tangent that the
closure already provides at the final load step. The price is a small ripple of the objective as cuts and stitches
cross the elements, and a cut band of finite width. A closure takes about 45\,s on the 0.5\,mm grid, so a design
problem of up to eight variables and 32 constraints is solved in about an hour with two threads on a personal
computer. Since one adjoint solve gives the gradient with respect to all
variables, more design variables do not raise the cost of an iteration.

Natural extensions include anisotropic skin, with fibers along the lines of tension, the size and shape of the lesion,
the closure map itself, other local flaps such as the bilobed flap and the Z-plasty, and the uncertainty of the skin properties
\citep{Stowers2021}. As a final remark, the formulation applies to any local flap whose closure map can be written as
pairs of lips, since the design enters only through the level-sets and the stitch points.

\section{Conclusions}\label{sec:conclusions}

The rhombic transposition flap and its excision were optimized on a fixed grid with up to eight design variables and
32 constraints, with the skin modeled as a pre-stretched, wrinkling Ogden membrane. The model reproduces the strain
energies of the 48-design study of \citet{Spagnoli2022} with a median deviation of 5\,\%. The optimal flap angles,
$(-23^\circ,31^\circ)$, describe a narrow flap turned towards its pivot that adds 63\,\% less energy than the Limberg
flap and 15\,\% less than the best swept design, at the same suture-line stress. Free cut lengths
lower the energy by a further 9\,\%, mainly through a longer second cut, and fitting the excision to the lesion and
its margin by 35\,\%. With all eight variables, the energy is 78\,\% below that of the Limberg flap, and 59\,\% below that of a Limberg
flap with its excision fitted to the lesion margin. A finer grid and a narrower cut
band lower all energies, by about 2\,\% and a fifth, but keep the optimized excision and the gains against the
Limberg flap. A wider undermining keeps the optimum as well, but reduces its gain against the Limberg flap from 78 to
52\,\% at $R=300$\,mm. Under an unequal in-vivo tension, the optimal flap turns with the direction of the tension, as
expected. The suture-line stress, measured as the net traction over the spacing of the sutures, can be bounded by the
$p$-mean. With all eight variables, lowering it by 41\,\% against the energy optimum raises the energy by only
2.3\,\%. This is cheap because the energy is nearly flat along $\theta_1$, whereas the suture-line stress is not.

\section*{CRediT authorship contribution statement}
Niels Aage: Conceptualization, Methodology, Software, Writing -- original draft, Supervision. Brandon M. Parks:
Conceptualization, Methodology, Writing -- review \& editing.

\section*{Declaration of competing interest}
The authors declare that they have no known competing financial interests or personal relationships that could
have appeared to influence the work reported in this paper.

\section*{Data availability}
The code used can be made available upon reasonable request.

\section*{Declaration of generative AI and AI-assisted technologies in the manuscript preparation process}
During the preparation of this work the authors used Claude (Anthropic) in order to write code for the plots and to edit the text. The authors
reviewed and edited the content as needed and take full responsibility for all content of the article.

\section*{Acknowledgements}
This research is funded wholly by Independent Research Fund Denmark under Grant ID 4286-00207B for the OptCut
Project.

\appendix
\section{Membrane stresses and tangent}\label{app:membrane}
With the in-plane right Cauchy-Green tensor $\bC=\bF^{\top}\bF$, its eigenvalues $c_1\ge c_2$, the squared principal
stretches, and $\hat\alpha=\alpha_1/2$, the Ogden energy of Eq.~(\ref{eq:ogden}) is
\begin{equation}
  w(c_1,c_2) = \frac{\mu_1}{\alpha_1}\bigl(c_1^{\hat\alpha} + c_2^{\hat\alpha} + (c_1c_2)^{-\hat\alpha} - 3\bigr),
\end{equation}
and the relaxed energy is
\begin{equation}
  \cW_{\mathrm O} = \begin{cases}
  w(c_1,c_2), & c_1c_2^2\ge1,\\
  w\bigl(c_1,c_1^{-1/2}\bigr), & c_1>1,\ c_1c_2^2<1,\\
  0, & c_1\le1,
  \end{cases}
  \label{eq:relax}
\end{equation}
for the taut, wrinkled and slack states. The taut condition reads $\lambda_2\ge\lambda_1^{-1/2}$: the second principal
stretch is at least the width of an incompressible strip in uniaxial tension $\lambda_1$. The energy is evaluated
through the invariants $\mathrm{tr}\,\bC$, $\det\bC$ and $(c_1-c_2)^2$, which are polynomials in $\bC$, so that the
energy, the stresses and the tangent stay smooth where $c_1=c_2$, as in the equibiaxial in-vivo state. In the taut
state, the second Piola-Kirchhoff stress $\bS$ follows from the derivatives of $w$ with respect to these invariants. In
the wrinkled state, with $f(c_1)=w(c_1,c_1^{-1/2})$, it is a uniaxial stress along the first principal direction
$\mathbf N_1$,
\begin{equation}
  \bS = 2f'(c_1)\,\mathbf N_1\otimes\mathbf N_1,
\end{equation}
and in the slack state it is zero. The regularization in Eq.~(\ref{eq:ogden}) adds its neo-Hookean stress in every
state. The first Piola-Kirchhoff stress is $\bP=\bF\bS$, and the tangent $\mathsf A=\partial\bP/\partial\bF$ is computed
by complex-step differentiation of $\bP$ \citep{SquireTrapp1998}. With the in-vivo stretch, the energy and
the stress with respect to $\bF_n$ are
\begin{equation}
  W(\bF_n) = \cW(\bF_n\bF_0),\qquad \bP_n = \bP(\bF_n\bF_0)\,\bF_0^{\top},
\end{equation}
and the tangent transforms accordingly. As the membrane is incompressible, $W$ is the energy per unit volume of the
in-vivo tissue. The element energy $\Psi_{T,e}$, residual $\br_{T,e}$ and tangent $\bK_{T,e}$ are the standard $2\times2$
Gauss sums of the bilinear element, multiplied by the in-vivo thickness $t$.

\section{Sutures: assembly, quadrature and stiffness}\label{app:sutures}
Spring $j$ joins the stitch points $\bX_j^+$ and $\bX_j^-$, which lie in host elements with the nodes $\bx_a^+$ and
$\bx_a^-$, $a=1,\dots,4$. Its gap is
\begin{equation}
  \mathbf t_j = \sum_{a=1}^4 N_a(\bX_j^+)\,\bx_a^+ - \sum_{a=1}^4 N_a(\bX_j^-)\,\bx_a^- = \bigl[\bD\bx\bigr]_j ,
\end{equation}
where $\bD$ has two rows per spring, one per displacement component, with $N_a(\bX_j^+)$ and $-N_a(\bX_j^-)$ in the
columns of these nodes. With $\mathbf A$ the diagonal matrix of the $\alpha_j$, each twice, the energy of the springs,
their forces on the nodes and their stiffness are
\begin{equation}
  \begin{aligned}
  \Pi_S &= \frac{\Lambda}{2}\,\bx^{\top}\bD^{\top}\mathbf A\bD\,\bx,\\
  \br_S &= \Lambda\,\bD^{\top}\mathbf A\bD\,\bx,\qquad \bK_S = \Lambda\,\bD^{\top}\mathbf A\bD .
  \end{aligned}
  \label{eq:springs}
\end{equation}
As the shape functions sum to one, every row of $\bD$ sums to zero, so a rigid translation leaves the gaps unchanged.
$\bK_S$ does not depend on $\bu$, and it is symmetric and positive semidefinite.

With $\alpha_j = \kappa_s l_0/n_s$, the springs of a pair are the midpoint rule, with $n_s$ points, for a penalty on
the gap along the lips,
\begin{equation}
  \frac{\Lambda}{2}\sum_j\alpha_j\,|\mathbf t_j|^2 \approx \frac{\Lambda}{2}\,\kappa_s l_0\int_0^1|\mathbf t(s)|^2\,\dd s ,
  \label{eq:linepenalty}
\end{equation}
where $s$ is the fraction of arc length at which the two lips are joined, and $l_0\,\dd s$ is the arc length on a lip
of the Limberg length. The integrand has a kink wherever one of the lips crosses an element edge, and the points do
not follow these kinks, so the rule is not exact. For a straight lip shifted along itself over the grid, the energy of
a smooth gap field is integrated with errors of up to 0.7\,\% with one stitch per element length and 0.02\,\% with
four.

As $\kappa_s\to\infty$, the springs enforce $\mathbf t_j=\mathbf 0$ at four points per element length on two lips whose
bilinear traces have their kinks at different places, so the energy keeps rising with $\kappa_s$
(Table~\ref{tab:kappa}). One stitch per element length, or one spring on the mean gap of each group of four
consecutive stitches, gives one constraint per element length instead. For both, the energy converges as
$1/\kappa_s$, and the two limits differ by 1.6\,\% on the 1\,mm grid and by 0.2\,\% on the 0.5\,mm grid. With
$\kappa_s = 10^4\,Et$, the energy of the Limberg flap, which needs the largest closing forces, is 0.9\,\% below the
mean of the two limits on the 0.5\,mm grid. Under the unequal stretch of Section~\ref{sec:orientation}, the same
$\kappa_s$ leaves gaps of at most 0.38\,mm in the Limberg flap, against 0.29\,mm under the equibiaxial stretch.

\begin{table*}[h]
\centering
\caption{Energy added by the surgery, $\Delta W$ (mJ), of the closed Limberg flap against the line stiffness
$\kappa_s$ of the sutures, for three closures on the grids with $h=1$ and 0.5\,mm. In parentheses: the largest gap
(mm) of the closure with a spring on every stitch.}
\label{tab:kappa}
\small
\begin{tabular}{L{6.2cm}lllll}
\toprule
Closure & $h$ (mm) & $10^3\,Et$ & $10^4\,Et$ & $10^5\,Et$ & $10^6\,Et$ \\
\midrule
Spring on every stitch, four per element length & 1 & 52.62 (0.92) & 56.64 (0.30) & 59.80 (0.08) & 62.23 (0.02) \\
 & 0.5 & 52.39 (0.93) & 56.07 (0.29) & 57.96 (0.08) & 59.25 (0.02) \\
One stitch per element length & 1 & & 54.76 & 55.10 & 55.14 \\
 & 0.5 & & 55.84 & 56.47 & 56.54 \\
Mean gap of four stitches & 1 & & 55.42 & 55.92 & 55.99 \\
 & 0.5 & & 55.89 & 56.59 & 56.67 \\
\bottomrule
\end{tabular}
\end{table*}

The tightening starts at $\Lambda_0=10^{-3}$, halved until Newton's method converges, and continues with
\begin{equation}
  \begin{aligned}
  \Lambda_n &= \min(\gamma_k\Lambda_{n-1},\,1),\\
  \gamma_k &= 1 + 0.5^{k-1}(\gamma_0-1),\qquad \gamma_0 = 1.5,
  \end{aligned}
\end{equation}
where $k$ grows by one whenever Newton's method fails, and the increment is then retried from the last converged
state. The closed state does not depend on the path: a slower path from $\Lambda=10^{-7}$ ends at the same energy
within $10^{-10}$.

\section{Model parameters and box constraints}\label{app:parameters}
\begin{table*}[h]
\centering
\caption{Parameters of the default problem (Section~\ref{sec:results}); the settings of single examples are
stated where they are used.}
\label{tab:params}
\small
\begin{tabular}{L{0.36\textwidth}L{0.58\textwidth}}
\toprule
Parameter & Value \\
\midrule
Skin disc & radius $R=100$\,mm, centered on the lesion, fixed edge; in-vivo thickness $t=1.5$\,mm \\
Ogden model & $\mu_1 = 110$\,Pa, $\alpha_1 = 26$, i.e.\ $\mu_0 = 1.43$\,kPa, $E = 3\mu_0$ \citep{Spagnoli2022} \\
Stabilization of wrinkled and slack skin & $\eta = 0.01$ \\
In-vivo stretch & $\lambda_0 = 1.1$, equibiaxial; Section~\ref{sec:orientation}: $\lambda_a = 1.15$ along the angle $a$, $\lambda_b = 1.05$ across \\
Structured box, element size & $[-38,44]\times[-64,24]$\,mm, $h = 0.5$\,mm \\
Ring & 14 layers, growth 1.25 \\
Cut band, smoothing & $\delta = 3$\,mm, $\varepsilon = 1$\,mm \\
Void material & weak linear, $E_{\mathrm{void}} = 10^{-6}E$ \\
Sutures & $\kappa_s = 10^4\,Et$ per mm, $n_s = 4l_0/h$ stitches per lip, $l_0=30$\,mm \\
Lesion, margin & $r_L = 8$\,mm, centered at the origin, $m_L = 3$\,mm \\
Flap base, scar length, box margin, neck & $\ge 15$\,mm, $\le 120$\,mm, 5\,mm, $\ge 4$\,mm \\
MMA & move limit 0.03, objective scale 100, $c=10^5$, stop at a change of $10^{-3}$ or 100 iterations (P5: 200) \\
Tightening, Newton & $\Lambda$ from $10^{-3}$ to 1, relative residual $10^{-9}$ \\
\bottomrule
\end{tabular}
\end{table*}

The box constraints keep every vertex $V\in\{A,\dots,F\}$ at least $b=5$\,mm inside the structured box
$[x_{\min},x_{\max}]\times[y_{\min},y_{\max}]$:
\begin{equation}
  \begin{aligned}
  &\frac{x_{\min}+b-V_x}{l_0}\le0,\qquad \frac{V_x-x_{\max}+b}{l_0}\le0,\\
  &\frac{y_{\min}+b-V_y}{l_0}\le0,\qquad \frac{V_y-y_{\max}+b}{l_0}\le0 .
  \end{aligned}
\end{equation}

\section{Sensitivities of the residual and of the suture-line stress}\label{app:sensitivities}
The right-hand side of the adjoint problem of Section~\ref{sec:sensitivities}, $-\partial\Delta W/\partial\bu$, is the negative
internal force of the skin and the partly void elements, and the explicit derivative of the objective is that of the
void fractions,
\begin{equation}
  \frac{\partial\Delta W}{\partial z_k} = \sum_e\frac{\partial\Delta W}{\partial\chi_e}\,\frac{\partial\chi_e}{\partial z_k} .
\end{equation}
By Eq.~(\ref{eq:chi}), with $H_E = H(-\phi_E;\varepsilon)$, $H_I = H(-\phi_I;\varepsilon)$ and the derivative $H'$
of $H$,
\begin{equation}
    \begin{aligned}
  \frac{\partial\chi_e}{\partial z_k} = {}&-(1-H_I)\,H'(-\phi_E)\,\frac{\partial\phi_E}{\partial z_k}\\
  &-(1-H_E)\,H'(-\phi_I)\,\frac{\partial\phi_I}{\partial z_k} .
  \end{aligned}
\end{equation}
The distance to a segment $PQ$, the closest side of the rhombus or the closest segment of the incisions, moves with
its end points at the fixed parameter $t^*$ of the closest point. With the unit vector $\hat{\mathbf n}$ from the
closest point to the element centroid,
\begin{equation}
  \frac{\partial\phi}{\partial z_k} = -\hat{\mathbf n}\cdot\Bigl((1-t^*)\,\frac{\partial P}{\partial z_k}
  + t^*\,\frac{\partial Q}{\partial z_k}\Bigr),
\end{equation}
times the sign of $\phi_E$ for the excision. With $\gamma_e = 1-\chi_e$, $e_v = E_{\mathrm{void}}/E$, the residual
$\br_{T,e}$ and the tangent $\bK_{T,e}$ of the skin element, and the stiffness matrix $\bK_L$ of the linear element,
$\Psi_{L,e} = \tfrac12\bu_e^{\top}\bK_L\bu_e$, the interpolation~(\ref{eq:interp}) gives
\begin{equation}
  \begin{aligned}
    \frac{\partial W_{\mathrm{tot}}}{\partial\chi_e} &= -\Psi_{T,e}(\gamma_e\bu_e)
  - \gamma_e\,\br_{T,e}(\gamma_e\bu_e)\cdot\bu_e\\
  &\quad + e_v\,\Psi_{L,e}(\bu_e),\\
  \frac{\partial W_{\mathrm{ref}}}{\partial\chi_e} &= -W(\mathbf I)\,t\,A_e,\\
  \frac{\partial\br_e}{\partial\chi_e} &= -2\gamma_e\,\br_{T,e}(\gamma_e\bu_e)
  - \gamma_e^2\,\bK_{T,e}(\gamma_e\bu_e)\,\bu_e + e_v\,\bK_L\bu_e .
  \end{aligned}
\end{equation}
The design thus enters the residual through the void fractions and through the stitch points,
\begin{equation}
  \begin{aligned}
  \bet^{\top}\frac{\partial\br}{\partial z_k} ={}& \sum_e \bet_e^{\top}\frac{\partial\br_e}{\partial\chi_e}
  \frac{\partial\chi_e}{\partial z_k}
  + \sum_j\alpha_j\bigl[\boldsymbol\Lambda_j\cdot(\bF_n^+\boldsymbol\xi^+_{jk} - \bF_n^-\boldsymbol\xi^-_{jk})\\
  &+ \mathbf t_j\cdot(\mathbf L^+\boldsymbol\xi^+_{jk} - \mathbf L^-\boldsymbol\xi^-_{jk})\bigr],
  \end{aligned}
\end{equation}
where $\boldsymbol\xi^\pm_{jk}=\partial\bX^\pm_j/\partial z_k$, $\boldsymbol\Lambda_j=\bet(\bX^+_j)-\bet(\bX^-_j)$,
and $\bF_n=\mathbf I+\nabla\bu$ and $\mathbf L=\nabla\bet$ are evaluated at the stitch points from the shape-function
gradients of their host elements. A point that moves by $\boldsymbol\xi$ at fixed nodal displacements moves its
current position by $\bF_n\boldsymbol\xi$ and its interpolated adjoint by $\mathbf L\boldsymbol\xi$. The derivatives
$\boldsymbol\xi^\pm_{jk}$ of the stitch points are computed by complex step through the lip geometry. Where a stitch
point crosses an element edge, the shape-function gradients jump, so the objective is only piecewise smooth in the
design.

The window means of the suture-line stress depend on the displacements and, through the current positions of the
stitch points, on the design. With $\zeta_k$ frozen within an iteration, the constraint~(\ref{eq:pmean}) changes with
the window means as
\begin{equation}
  \frac{\partial\sigma_{\mathrm{PM}}}{\partial s_i} = \frac{1}{4n_s}\Bigl(\frac{\max(s_i,0)}{\sigma_{\mathrm{PM}}}\Bigr)^{p-1},
\end{equation}
and the chain rule through the windows and the stitch stresses gives the partial derivatives with respect to the
nodal displacements, $\bar\bu$, and with respect to the design at fixed displacements, $\bar\bz$. One more adjoint
solve with the same factorization gives
\begin{equation}
  \bK\boldsymbol\mu = -\bar\bu,\qquad
  \frac{\dd g_7}{\dd\bz} = \frac{\zeta_k}{f\bar\sigma}\Bigl(\bar\bz + \boldsymbol\mu^{\top}\frac{\partial\br}{\partial\bz}\Bigr) .
\end{equation}
The gradients agree with central finite differences of re-solved closed states to $2\cdot10^{-9}$ to
$3\cdot10^{-5}$ for the objective and $3\cdot10^{-8}$ to $4\cdot10^{-6}$ for the stress constraint, except where a kink
of the objective lies inside the finite-difference stencil.

\bibliographystyle{cas-model2-names}
\bibliography{refs}

\end{document}